\documentclass[preprint,12pt]{elsarticle}
\usepackage{amsmath,amssymb,graphicx,geometry,bm,textcomp,siunitx}
\usepackage[utf8]{inputenc}
\usepackage{multirow}

\journal{Solar Energy Materials and Solar Cells}

\begin{document}

\begin{frontmatter}

\title{Technological guidelines for the design of tandem III-V nanowire on Si solar cells from opto-electrical simulations}

\author[IMEP]{Vladimir Maryasin\corref{cor1}}
\ead{vmariassin@gmail.com}
\cortext[cor1]{Corresponding author}
\author[IMEP]{Davide Bucci}
\author[IMEP]{Quentin Rafhay}
\author[IMEP]{Federico Panicco}
\author[IMEP]{J\'{e}r\^{o}me Michallon}
\author[IMEP]{Anne Kaminski-Cachopo}
\address[IMEP]{Univ. Grenoble Alpes, CNRS, Grenoble INP \fnref{IOE}, IMEP-LAHC, 3 Parvis Louis Neel, F-38000 Grenoble, France\fnref{label3}}
\fntext[IOE]{Institute of Engineering Univ. Grenoble Alpes}


\begin{abstract}
Effect of geometrical and structural parameters on the efficiency of the tandem solar cell based on the III-V nanowire array on silicon is studied by the means of coupled opto-electrical simulations.
A close to realistic structure, consisting of AlGaAs core-shell nanowire array, connected through a tunnel diode to a Si subcell is modelled, revealing the impact of top contact layer, growth mask and tunnel junction.
Optical simulation of the tandem structure under current matching condition determine optimal geometrical parameters of the nanowire array.
They are then used in the extensive electrical optimization of the radial junction in the nanowire subcell. 
Device simulations show the necessity of high doping of the junction in order to avoid full shell depletion. 
The influence of bulk and surface recombination on the performance of the top subcell is studied, exposing the importance of the good surface passivation near the depleted region of the radial $p-n$ junction.
Finally, simulations of the fully optimized tandem structure show that a promising efficiency of $\eta = 27.6\%$ with the short-circuit current of $J_\mathrm{SC} = \SI[per-mode=symbol]{17.1}{\milli\ampere\per\centi\metre\squared}$ can be achieved with reasonable bulk and surface carrier lifetime.
\end{abstract}


\end{frontmatter}


\section{Introduction}

Semiconductor nanowires (NW) have recently emerged as promising candidates for a new generation of photovoltaic devices \cite{Garnett11}, since their excellent antireflection and light-trapping characteristics are superior to those of planar cells \cite{Kupec10, Wallentin13}.
In addition, multijunction solar cells present a way to overcome the fundamental Shockley-Queisser efficiency limit, with the current efficiency-record-holding cell being a four junction cell under concentrated sunlight \cite{Green17}.
The outstanding performance of such cells is obtained thanks to selective absorption of different parts of the solar spectrum and minimization of energy losses from thermalization of photogenerated carriers.
Combining the benefits of the two structures, III-V nanowires (NW) on silicon tandem cell could be hence considered as potentially one of the most efficient structures within dual-junction solar cells.

The choice of silicon as one of the active layers for a tandem cell is dictated by the element abundance and technological maturity.
III-V materials in return have outstanding light absorption properties \cite{Beer66}. Moreover, the use of ternary alloys of III-V materials gives the possibility to continuously fine-tune the band gap of the top cell material to obtain the best possible performance with the given substrate.
Particularly, $E_g = \SI{1.7}{\electronvolt}$ is known to provide the highest ultimate efficiency with silicon in a tandem cell \cite{Kurtz90}.

In addition to their excellent light absorption properties, nanowires have small substrate-contact area. This leads to a better strain relaxation from the strong lattice mismatches and the formation of antiphase boundaries on the III-V/Si interface \cite{Kavanagh10}.
Finally, the use of nanowires introduces another promising concept for the solar cell application, i.e. the radial $p-n$ junction. Indeed, core-shell structure benefits from decoupled effective optical and electrical lengths, allowing to maximize the former and minimize the latter at the same time \cite{Kayes05}.
Therefore, III-V nanowires allow to significantly reduce material needs without compromising absorption or performance \cite{LaPierre13}.

The complexity of the proposed tandem system, however, implies that an extensive number of parameters influences the performance of the solar cell. Some of them, such as carrier lifetime, cannot be precisely controlled in real devices.
Other parameters have a non-monotonic impact on the efficiency, so that the system has to be carefully tailored in order to achieve the highest efficiency. 
LaPierre in his pioneering theoretical works \cite{LaPierre11, LaPierre11a} presented a first set of design rules for the perfectly absorbing III-V nanowire on silicon tandem cell.
Then, a number of studies with more sophisticated coupled opto-electronic simulations of the system followed \cite{Huang12,Bu13,Wang15}, exploring various aspects of cell design.

In this work, state-of-the-art numerical techniques are employed to improve the understanding of cell performance dependencies on various internal and external parameters.
The main goal of this study hence consists in providing useful and practical guidelines for the complex technological optimization of the III-V nanowire on silicon tandem structure.
In this work, $\mathrm{Al}_{0.2}\mathrm{Ga}_{0.8}\mathrm{As}$ $(E_g = \SI{1.7}{\electronvolt})$ is chosen as a top cell material, but the main results are expected to be similar for other alloys with the same band gap, such as $\mathrm{Ga}\mathrm{As}_{0.8}\mathrm{P}_{0.2}$ or $\mathrm{Ga}_{0.35}\mathrm{In}_{0.65}\mathrm{P}$.


From the modelling perspective, the most rigorous approach would imply the coupling of optical and electrical simulation, so that the full characteristic of every possible geometry is simulated, considering each time varying parasitic effects. 
This approach is however unrealistic from a computational perspective, so an alternative two-step procedure is employed in this work. 
First, optical simulations are performed by the home-built rigorous coupled wave analysis (RCWA) solver, which is the main methodological feature of this work.
The efficiency of our optical tool allows the sampling of the multidimensional parameter space for the self-consistent simulation of a whole complex 3D structure close to the realistic one. 
Furthermore, in contrast with the Beer-Lambert approximation used in all preceding works for the absorption in the bottom cell, our simulations, are based on the calculation of the spatially-resolved photogeneration rate in silicon. 
The coupling with electrical simulations of the tandem cell is carried out afterwards using the TCAD commercial simulator Sentaurus, which accounts for a wide variety of electronics effects. 
Thus, the study of the auxiliary layers effects (top contact, passivation, growth mask) on the current photogenerated in the dual-junction system is reported.

In the first part of the work, optical simulations are carried out to optimize light absorption over the geometrical parameters of the tandem cell under the current-matching condition.
In the second part of the paper, charge transport mechanisms are studied and carrier collection efficiency is optimized by the means of electrical device modelling coupled with the optical simulations. 
Core-shell design of the nanowire subcell requires careful and precise selection of the $p-n$ junction parameters due to the possibility of complete depletion of either core or shell \cite{Chia13, Li15b}.
Therefore, a broad numerical study is performed and design rules for radial junctions in the nanowires are proposed. 
After that, the influence of bulk and surface Shockley-Read-Hall recombination on the junction performance is investigated.
Finally, electrical simulations of the fully optimized solar cell consisting of two junctions connected by a tunnel diode are performed. 
The influence of the NW-Si contact scheme on the overall cell performance is finally discussed.


\section{Optical simulations}
\subsection{Model}

\begin{figure}[ht]
\centerline{
\includegraphics[width=0.21\columnwidth]{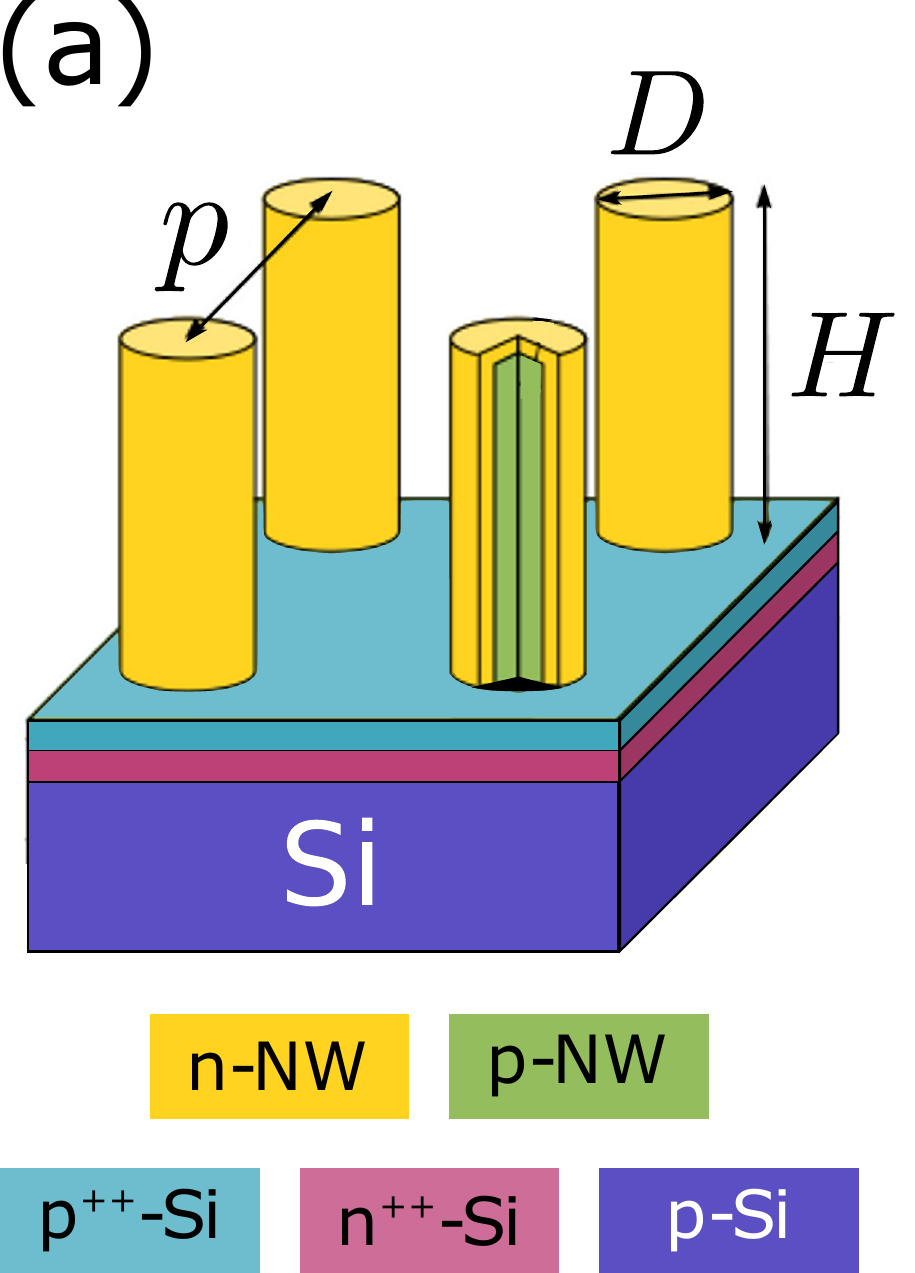} \hspace{7mm}
\includegraphics[width=0.18\columnwidth]{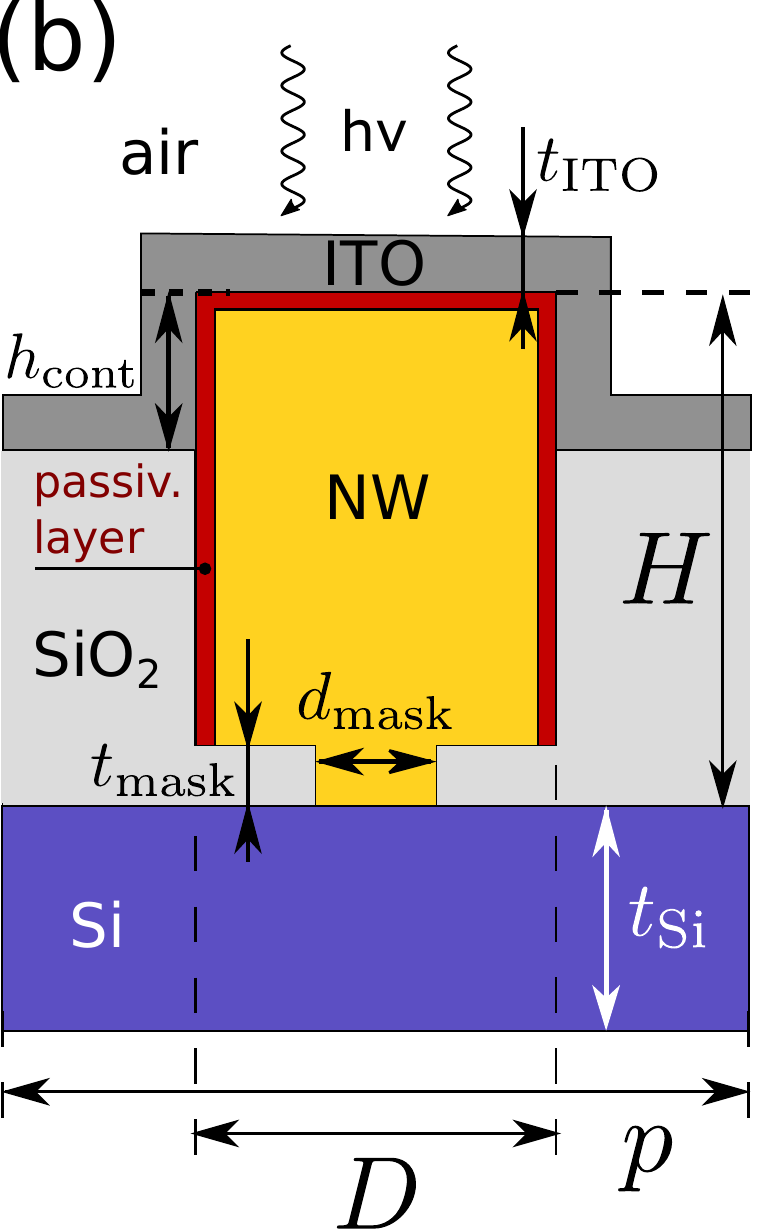}
}
\caption{(a) Schematic of AlGaAs core-shell nanowire on silicon tandem solar cell. (b) Cross-section view of one period of the structure with geometrical and structural parameters of the studied tandem cell, relevant for optical simulations. 
}
\label{fig:scheme}
\end{figure}

The first parameters to be targeted from the technological point of view are the periodicity of the nanowires $p$, their diameter $D$ and height $H$. 
Due to the sub-wavelength structure of the ordered NW array, it is possible to select these geometrical parameters in order to benefit from resonant light trapping.
So the first step of the work consists in the light absorption optimization in the tandem cell. 
Numerical solution of Maxwell's equations in the system is carried out by the home-built RCWA solver \cite{Moharam95, Bucci12}. This state-of-the-art optical simulation tool enables the study of light, absorbed by complex 3D multilayer structures of arbitrary length, in a reasonable computational time. 
Our study follows the simulation strategy and choice of parameters, described in detail by Michallon \textit{et al.} \cite{Michallon14}.

The investigated system, which replicates a realistic tandem cell, is presented in Fig.~\ref{fig:scheme} (b), with structural parameters summarized in table \ref{tab:opt}.
The top subcell of the studied tandem features a periodic array of $\mathrm{Al}_{0.2}\mathrm{Ga}_{0.8}\mathrm{As}$ cylindrical nanowires.
Their top and side surface is covered with a $t_\mathrm{pass} = \SI{5}{\nano\metre}$ layer of  $\mathrm{Al}_{0.5}\mathrm{Ga}_{0.5}\mathrm{As}$.
As was shown by Songmuang \textit{et al.}, a few nanometres of $\mathrm{Al}_{0.33}\mathrm{Ga}_{0.67}\mathrm{As}$ can efficiently suppress surface recombination of $\mathrm{Ga}\mathrm{As}$ nanowires \cite{Songmuang16}. An alloy with the same band gap difference is therefore chosen as a passivation layer in our study.
Nanowires are then encapsulated in $\mathrm{SiO}_2$ for mechanical and chemical stability of the device.
An additional \SI{10}{\nano\metre} layer of silica serves as a growth and isolation mask as well as a passivation layer for the bottom subcell. 
A $t_\mathrm{ITO} = \SI{100}{\nano\metre}$ layer of indium-tin oxide (ITO) is used as a transparent contact, which covers the top and a fraction of the side of the nanowire, labelled $h_\mathrm{cont}$.
Such thickness is a good trade-off between the optical and electrical properties. Indeed, thinner ITO contacts have reduced conductivity \cite{Utsumi98}, while thicker ITO layers increase optical losses as nonvanishing absorption coefficient of ITO was taken into account in simulations \cite{Konig14}.

The studied bottom subcell consists of the $\mathrm{Si}$ substrate with a typical thickness of \SI{200}{\micro\metre}. In the optical simulations it is modelled by a semi-infinite silicon slab with absorption calculated in the first \SI{200}{\micro\metre}. Simulations of such system can be conveniently done with RCWA, reflections from the absent back silicon/air interface are not present and cannot yield unphysical interference effects induced by the simulated coherent light.
This approach correctly takes into account the diffracted light induced by the nanowire array in contrast with the Beer-Lambert approximation, commonly used for the substrate. 
However, similarly to the  Beer-Lambert approach, our method may somewhat underestimate light absorption in Si due to the absence of back reflection. 
This point is further discussed among the loss analysis at the end of section \ref{sec:OptRes}.
The whole structure is illuminated from the air on top by monochromatic planewaves. 
Optical indices of all the above mentioned materials are taken from the Refractive index database \cite{Refractiveindex16}.

Geometrical parameters of the array: NW length $H$, pitch size $p$, and filling ratio $D/p$ are varied in order to increase photo-generated current in the structure.
For each set of parameters, spectral absorption $A(\lambda)_\mathrm{NW/Si}$ in both subcells is calculated. Then, the ultimate photocurrent density, i.e. current density in the detailed balance limit, is obtained by integrating spectral absorption with the ASTM AM1.5G solar radiation density spectrum $I(\lambda)$ \cite{AM1.5D}:
\begin{equation}
J_\mathrm{NW/Si} = \frac{e}{hc}\int_{\SI{300}{\nano\metre}}^{\SI{1100}{\nano\metre}} A(\lambda)_\mathrm{NW/Si}I(\lambda)\lambda d\lambda.
\label{ultimatecurrent}
\end{equation}
Due to the series connection of the NW and Si photogenerating junctions, the total current flowing through the tandem cell is determined by the smaller of the two $J_ \mathrm{NW/Si}$. 
Thus, to avoid inefficient absorption in one of the subcells, the so-called current matching condition is imposed on the structure: $J_\mathrm{NW} = J_\mathrm{Si} = J_\mathrm{CM}$.
Therefore, both current densities have to be maximized simultaneously and the optimal configuration is found to be the one with the maximal matched current $J_\mathrm{CM}$.

\begin{table}[ht]
\footnotesize
\centering
\begin{tabular}{llc}
\hline
Parameter & Description & Value \\
\hline
&& \SI{0.9}{\micro\metre} -- \SI{2.4}{\micro\metre}  \\
$H$&height of the nanowire& \SI{1.5}{\micro\metre} \\
&& \SI{150}{\nano\metre} -- \SI{850}{\nano\metre}  \\
$p$&nanowire array pitch size& \SI{550}{\nano\metre} \\
&& \SI{30}{\nano\metre} -- \SI{765}{\nano\metre} \\
$D$&diameter of the nanowire& \SI{330}{\nano\metre} \\
\hline
$t_\mathrm{Si}$ & Si subcell thickness & \SI{200}{\micro\metre} \\
$t_\mathrm{ITO}$ & thickness of the ITO layer & \SI{100}{\nano\metre} \\
$t_\mathrm{pass}$ & thickness of the passivation layer & \SI{5}{\nano\metre} \\
$h_\mathrm{cont}$ & vertical extent of top contact & \SI{300}{\nano\metre} \\
$t_\mathrm{mask}$ & thickness of the $\mathrm{SiO_2}$ mask & \SI{10}{\nano\metre} \\
$d_\mathrm{mask}$ & hole diameter in the $\mathrm{SiO_2}$ mask  & \SI{50}{\nano\metre} \\
\hline
\end{tabular}
\caption{Geometrical parameters of the studied tandem cell. For $H$, $p$ and $D$ both variation range and optimal values are given.}
\label{tab:opt}
\end{table}


\subsection{Results of optical modelling}
\label{sec:OptRes}

\begin{figure*}[ht]
\centerline{
\includegraphics[width=0.4\columnwidth]{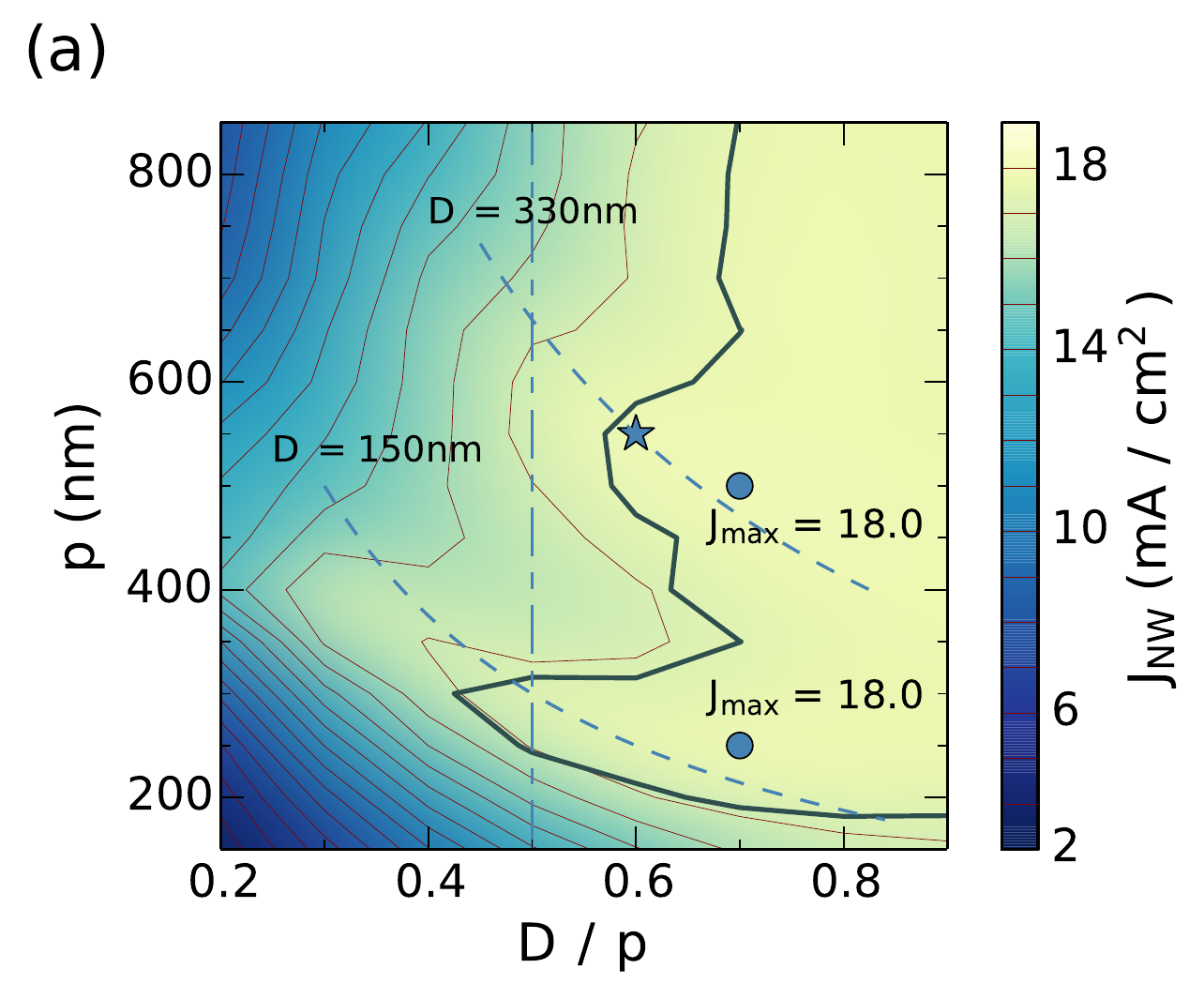} \hspace{10mm}
\includegraphics[width=0.4\columnwidth]{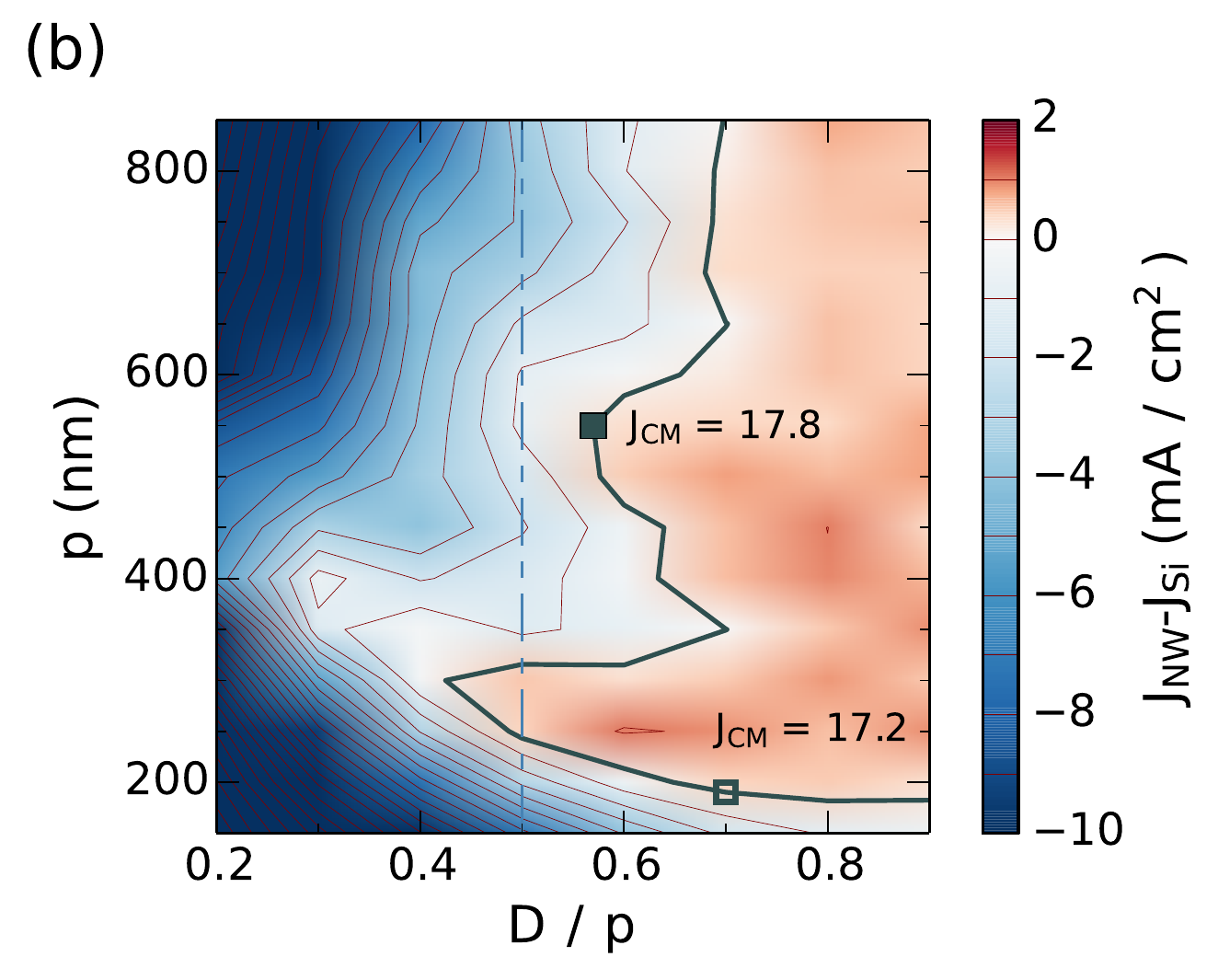}
}
\caption{(a) Map of ultimate photogenerated current in the top cell - \SI{1.5}{\micro\metre} high $\mathrm{Al}_{0.2}\mathrm{Ga}_{0.8}\mathrm{As}$ nanowire array.  
Circle markers denote two configurations with maximal ultimate NW photocurrent. Star marker corresponds to the geometry which is used for further electrical simulations $D = \SI{330}{\nano\metre}$, $p=\SI{550}{\nano\metre}$ with $J_\mathrm{NW} = \SI[per-mode=symbol]{17.8}{\milli\ampere\per\centi\metre\squared}$.
Dashed hyperbolic lines show configurations with fixed $D = \SI{150}{\nano\metre}$ and $D = \SI{330}{\nano\metre}$. Dash-dotted vertical line delimits high-current configurations with $D/p > 0.5$.
(b) Photocurrent difference in the top and bottom subcells defines current matched geometries, shown by the thick gray line on both panels. Square marker corresponds to the array configuration with the maximal matched current.}
\label{fig:opt_NW}
\end{figure*}

Our optical simulation results are presented in Fig.~\ref{fig:opt_NW} (a), which shows a color map of photocurrent density in the AlGaAs nanowire array for \SI{1.5}{\micro\metre} high nanowires as a function of the array period $p$ and filling factor $D/p$.
Two distinctive regions of the color map can be observed. For $D/p < 0.5$, photogenerated current grows strongly with the filling fraction, while for $D/p \gtrsim 0.5$, it tends to saturate with a broad maximum. Increase of reflection at higher filling ratios $D/p$ \cite{Sturmberg14} is less pronounced, compared to the previous optical simulation results \cite{Huang12,Sturmberg14,Huang12a}. It could be explained by the presence of the additional ITO layer, covering the NW array from the top and providing smoother refractive index variation from air to the nanowire array.

Furthermore, the photogenerated current depends strongly on the nanowire diameter. 
Dashed lines in Fig.~\ref{fig:opt_NW} (a) denote configurations with fixed $D = \SI{150}{\nano\metre}$ and $D = \SI{330}{\nano\metre}$ and reproduce the main absorption features on the contour plot.
The behavior can be explained by performing a modal analysis of the absorbed light \cite{Michallon14}.
Photocurrent in the NW array is maximized when diameter-dependent spectral absorption resonance is located close to the absorption edge of the NW material \cite{Sturmberg14,Anttu13}. 
Two maxima with equal current density of \SI[per-mode=symbol]{18.0}{\milli\ampere\per\centi\metre^2}, are found close to these equidiameter geometries. 
It should be noted that usually the smaller-diameter absorption peak is reported to be dominant \cite{Huang12,Sturmberg14, Anttu13}. 
However, in our simulations with different combinations of auxiliary layers and different nanowire heights,
the top larger-diameter maxima was always greater or equal with the bottom one within the precision of the simulations.

To find the configurations that satisfy current matching condition, the difference of the photogenerated currents in two subcells is plotted in Fig.~\ref{fig:opt_NW} (b).
The thick gray line indicates the geometries with matched current, the line is replicated on panel (a) to demonstrate the absolute values of $J_\mathrm{CM}$ obtained in all these configurations. 
The highest matched current, marked by a filled square on panel (b), is detected close to the $D = \SI{330}{\nano\metre}$ absorption peak. This justifies the choice of $p = \SI{550}{\nano\metre}$, $D/p = 0.6$ and $D = \SI{330}{\nano\metre}$ NW array geometry as optimal for our tandem cell, it is denoted by a star in Fig.~\ref{fig:opt_NW} (a). The corresponding photogenerated currents in two subcells are $J_\mathrm{NW} = \SI[per-mode=symbol]{17.8}{\milli\ampere\per\centi\metre\squared}$ and $J_\mathrm{Si} = \SI[per-mode=symbol]{17.6}{\milli\ampere\per\centi\metre\squared}$ respectively.

To facilitate further system improvement, it is important to analyze optical losses that are present in this geometry.
They are calculated in analogy with Eq.~(\ref{ultimatecurrent}) by introducing the corresponding spectral reflectance $R(\lambda)$ or transmittance $T(\lambda)$ instead of $A(\lambda)$.
Transmission losses in the infra-red part of the spectrum constitute the largest part, equal to \SI[per-mode=symbol]{3.5}{\milli\ampere\per\centi\metre^2}. They do not depend much on the nanowire array geometry and are due to low absorption in silicon at $\lambda \geq \SI{1000}{\nano\metre}$.
These losses can be somewhat reduced: increasing the substrate thickness up to \SI{400}{\micro\metre} or adding an ideal back reflector will increase photocurrent in the Si subcell by \SI[per-mode=symbol]{1.2}{\milli\ampere\per\centi\metre^2}. 
Furthermore, despite the fact that nanowire array exhibits efficient anti-reflecting properties compared to planar cells, \SI[per-mode=symbol]{2.6}{\milli\ampere\per\centi\metre^2} is lost due to light reflection in the optimal configuration. However, modest radial disorder inevitably present in real systems may contribute to antireflection properties by allowing the broadband coupling of the incident light \cite{Sturmberg12, Foldyna13, Fountaine14a}. 
In addition to this, nonuniform conical shape of the nanowires \cite{Fountaine14a, Duan16} in a similar way may reduce the reflection losses, although absorption and carrier collection properties of nanocones are still debated \cite{Li15}.
An important \SI[per-mode=symbol]{1.4}{\milli\ampere\per\centi\metre^2} is absorbed by the ITO layer primarily in the ultraviolet part of the spectrum, therefore, large $t_\mathrm{ITO}$ or $h_\mathrm{cont}$ should technologically be avoided. 
Finally, optical effect of the thin passivation layer is within the precision of the simulations, and is neglected in the following.

\begin{figure}[ht]
\centerline{
\includegraphics[width=0.45\columnwidth]{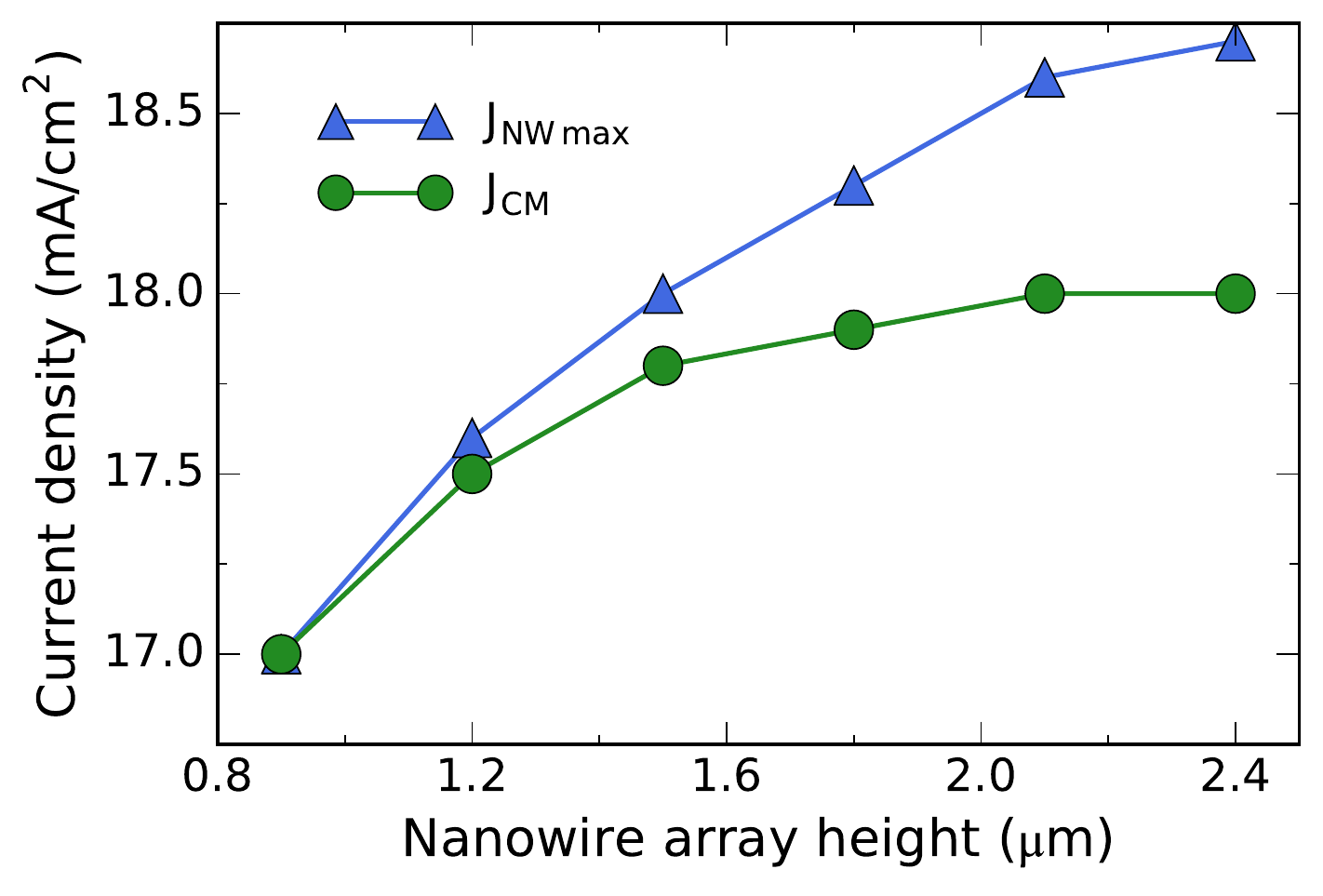}
}
\caption{Variation of top cell maximal photogenerated current and matched current flowing through the tandem cell with nanowire array height.}
\label{fig:H}
\end{figure}

\begin{figure}[ht]
\centering
\includegraphics[height=0.33\columnwidth]{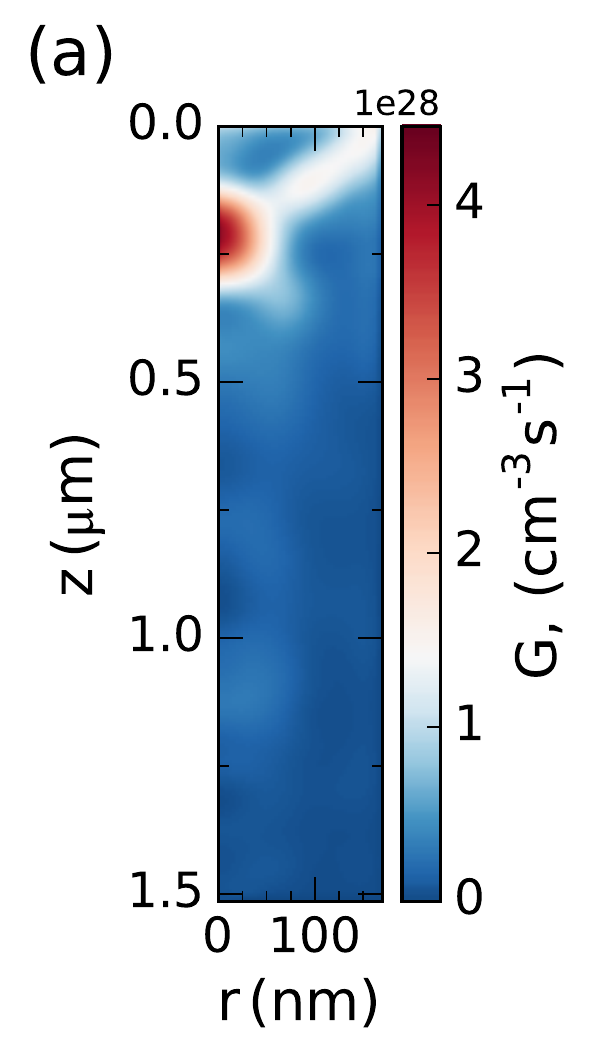} \hspace{10mm}
\includegraphics[height=0.33\columnwidth]{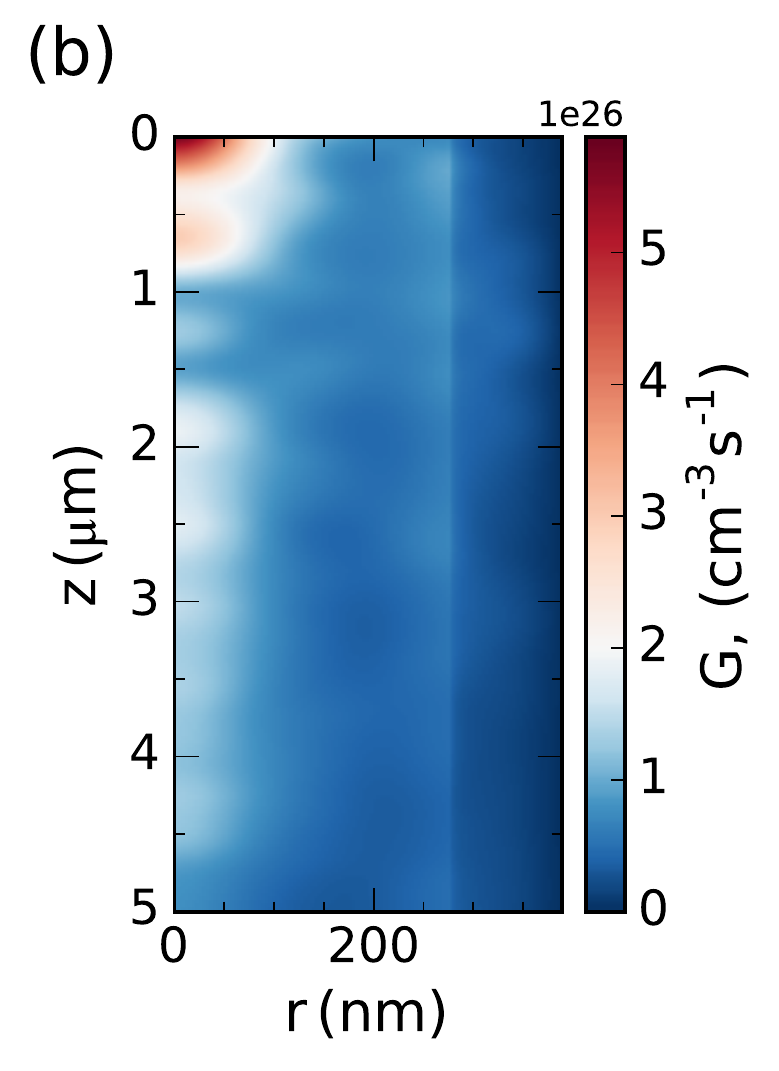}
\caption{(a) Spatially-resolved carrier generation rate in a single nanowire of the array with $D = \SI{330}{\nano\metre}$ and $H = \SI{1.5}{\micro\metre}$. Cylindrical coordinates are used with the averaging over the polar angle performed according to Eq.~(\ref{angle-averaging}). (b) Carrier generation rate in the top \SI{5}{\micro\metre} of the silicon substrate. 
}
\label{fig:GenRate}
\end{figure}

Following the same approach, tandem cell has been optimized for several nanowire array heights.
Figure \ref{fig:H} shows the evolution of the ultimate top cell photocurrent $J_\mathrm{NW}$ and matched current $J_\mathrm{CM}$ with varying $H$. 
First, it can be seen that the array of small nanowires with $H = \SI{0.8}{\micro\metre}$ can generate relatively high photocurrents.
In fact, the largest part of the light absorption happens within the first \SI{0.4}{\micro\metre} of the nanowire, as can be also observed from the carrier generation map in Fig.~\ref{fig:GenRate} (a). 
Further, maximal ultimate photocurrent continues to grow with increasing NW length as more red photons are absorbed by higher nanowires.
But more importantly, the matched photocurrent saturates for $H \geq \SI{1.5}{\micro\metre}$, so further increase of the NW height results only in the shift of the optimal geometries towards sparser arrays and slightly thinner nanowires \cite{Huang12a, Hu13}.
On the contrary, growing periodic arrays of regular vertical cylindrical nanowires with larger length is technologically very challenging. 
Therefore, the optimal nanowire length is fixed to \SI{1.5}{\micro\metre} for the rest of the paper.

\section{Electrical simulations}
\subsection{Model}
The ultimate photocurrent density calculated by equation (\ref{ultimatecurrent}) corresponds to the short-circuit current density under the assumption that every generated electron-hole pair gets collected. 
Current density of real devices is, however, considerably lower than the values in the idealized model of the optical simulations, in particular due to various recombination processes.
The aim of the second part of our study is thus to optimize carrier collection and study the impact of recombination mechanisms present in the tandem structure.

\begin{figure}[ht]
\centerline{
\includegraphics[width=0.3\columnwidth]{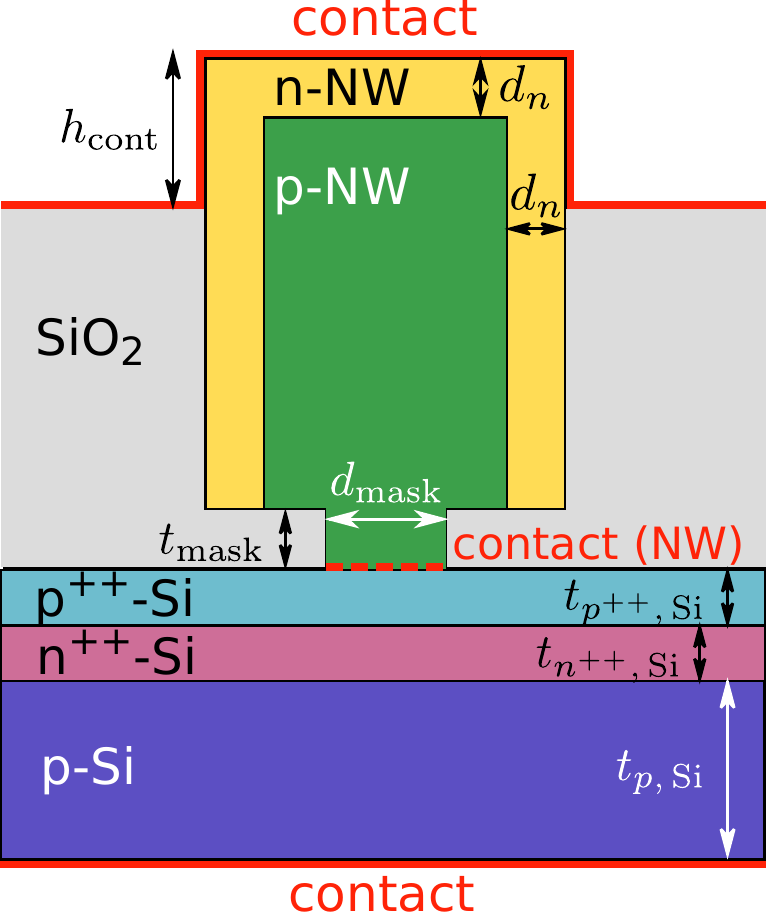}
}
\caption{Cross-section view of the period of the tandem cell showing the parameters, relevant for electrical simulations. Red dashed line shows the position of the bottom contact for the top cell modelling.}
\label{fig:el_system}
\end{figure}

The system under investigation is schematically shown in Fig.~\ref{fig:el_system} with optimal NW array parameters taken from section \ref{sec:OptRes} ($p = \SI{550}{\nano\metre}$, $D/p = 0.6$, $D = \SI{330}{\nano\metre}$, $H = \SI{1.5}{\micro\metre}$).
Nanowire with radial $p-n$ junction is connected in series with the planar $p - n^{++}$ silicon cell.  
A contact between the subcells is provided by a thin $p^{++} - n^{++}$ tunnel diode.

\begin{table*}[ht]
\footnotesize
\centering
\begin{tabular}{llc}
\hline
Parameter & Description & Value \\
\hline
&& \SI{15}{\nano\metre} -- \SI{50}{\nano\metre} \\
$d_n$&axial and radial shell thickness& \SI{20}{\nano\metre} \\
&& \SI{280}{\nano\metre} -- \SI{315}{\nano\metre} \\
$r_p$&core radius& \SI{310}{\nano\metre} \\
&& \SI{1e17}{\per\centi\metre\cubed} -- \SI{1e19}{\per\centi\metre\cubed} \\
$N = N_n = N_p$ & nanowire donor and acceptor doping concentration & \SI{5e18}{\per\centi\metre\cubed}  \\
\hline
$\mu_n$ & electron mobility in the nanowire  & \SI[per-mode=symbol]{1500}{\centi\metre\squared\per\volt\per\second} -- \SI[per-mode=symbol]{4500}{\centi\metre\squared\per\volt\per\second} \\
$\mu_p$ & hole mobility in the nanowire  & \SI[per-mode=symbol]{100}{\centi\metre\squared\per\volt\per\second} -- \SI[per-mode=symbol]{280}{\centi\metre\squared\per\volt\per\second} \\
$v_\mathrm{th}$ & thermal velocity of carriers  & \SI[per-mode=symbol]{1e7}{\centi\metre\per\second} \\
$\tau = \tau_n = \tau_p$ & SRH recombination lifetime  & \SI{1}{\nano\second} \\
$S = S_n = S_p$ & surface recombination velocity  & \SI[per-mode=symbol]{1e4}{\centi\metre\per\second} \\
$\sigma_n = \sigma_p$ & carrier  capture cross-section of trap states & \SI{1e-15}{\centi\metre\squared} \\
$N_\mathrm{TA} = N_\mathrm{TD}$ & density of surface trap states & \SI{1e12}{\per\centi\metre\squared} \\
$C_\mathrm{rad}$ & radiative recombination coefficient & \SI[per-mode=symbol]{1.8e-10}{\centi\metre\cubed\per\second} \\
$C_{\mathrm{Aug}}$ & electron and hole Auger recombination coefficients & \SI[per-mode=symbol]{1e-30}{\centi\metre^6\per\second} \\
\hline
$t_{p^{++}, \mathrm{Si}}$ & thickness of the tunnel junction layer &\SI{15}{\nano\metre} \\
$t_{n^{++}, \mathrm{Si}}$ & donor layer thickness of the silicon bottom cell &\SI{15}{\nano\metre} \\
$t_{p, \mathrm{Si}}$ & acceptor layer thickness of the silicon bottom cell &\SI{199.97}{\micro\metre} \\
$N_{p^{++}, \mathrm{Si}}$ & donor doping concentration in the tunnel junction & \SI{1e20}{\per\centi\metre\cubed} \\
$N_{n^{++}, \mathrm{Si}}$ & acceptor doping concentration in silicon & \SI{1e20}{\per\centi\metre\cubed} \\
$N_{p, \mathrm{Si}}$ & donor doping concentration in silicon & \SI{5e16}{\per\centi\metre\cubed} \\
\hline
\end{tabular}
\caption{Parameters of the electrical simulations of the tandem solar cell.  For $d_n$, $r_p$, and $N$ variation range and optimal values are given. $\mu_n$ and $\mu_p$ are doping-dependent mobilities, which vary with  $N$ according to Ref.~\cite{Adachi05}}
\label{tab:el}
\end{table*}

Continuity equations for electron and hole densities together with the Poisson's equation for the electrostatic potential are solved by the finite element method in Sentaurus TCAD \cite{Sentaurus}.
Charge accumulation in traps at the nanowire - $\mathrm{SiO_2}$ interface, which leads potentially detrimental band bending, is included in the simulations.
The traps are modelled by single level energy states located at the middle of the band gap.
Surface recombination velocities (SRV) for the electrons and holes $S_{n, p}$ are related to the corresponding surface density of donor and acceptor trap states $N_\mathrm{TA, TD}$, following:
\begin{equation}
S_{n, p} = \sigma_{n, p} v_\mathrm{th} N_\mathrm{TA, TD}
\label{SRV}
\end{equation}
with $\sigma_{n, p}$ denoting capture cross-section of the trap states and $v_\mathrm{th}$ denoting carrier thermal velocity. In this study, electrons and holes are assumed to have equal SRV $S = \SI[per-mode=symbol]{1e4}{\centi\metre\per\second}$. Similarly the densities of defect states are set to be equal $N_\mathrm{TD} = N_\mathrm{TA} = \SI[per-mode=symbol]{1e12}{\per\centi\metre\squared}$.

In addition to surface recombination, three bulk recombination mechanisms are taken into account: 
Shockley-Read-Hall (SRH) recombination through the midgap defect energy levels, radiative recombination of carriers and three-particle Auger recombination. The corresponding recombination coefficients for AlGaAs used in the current study are given in table \ref{tab:el}.

The spatially resolved carrier generation rate, calculated from the 3D electromagnetic field distribution, serves to couple optical and electrical simulations. Following Ref.~\cite{Michallon14}, it is defined as:
\begin{equation}
G(r,z) = \frac{1}{2 \hbar}\int \epsilon^{\prime\prime}(r,z,\lambda) \vert \mathbf{E}(r,z,\lambda)\vert^2 I(\lambda) d\lambda,
\label{GR}
\end{equation}
where $\epsilon^{\prime\prime}$ is the imaginary part of the medium permittivity and $I(\lambda)$ denotes solar irradiance spectrum. The electric field $\mathbf{E}(\mathbf{r},\lambda)$ is derived as a direct output of the optical simulations. To reduce the problem to 2D, averaging over the polar angle is used:
\begin{equation}
 \vert \mathbf{E}(r,z,\lambda)\vert^2 = \frac{1}{2\pi}\int\vert \mathbf{E}(r, \theta, z,\lambda)\vert^2d\theta.
\label{angle-averaging}
\end{equation}
In this work, generation rate calculation is employed for both top and bottom subcells. The latter is used to take into account nonuniformity of $G(r,z)$ and absorption enhancement in planar cells due to the nanowire coating \cite{Garnett10, Zhu10}.
Generation rate in a single nanowire and first \SI{5}{\micro\metre} of the $G(r,z)$ in silicon are presented in Fig.~\ref{fig:GenRate}.
Both panels illustrate strong spatial dependence of absorption with the main peak of $G(r,z)$ in silicon located directly below the nanowire.
It should be noted that an enlarged cylinder with the maximal radius $r_\mathrm{max} = p\sqrt{2}/2$ is used to describe the square unit cell of the substrate in cylindrical coordinates (see \ref{sec:App} for more details).

The non-local tunnelling model of Sentaurus TCAD \cite{Sentaurus} is activated to simulate the role of the $p^{++}/n^{++}$ tunnelling diode of the tandem structure. The band-to-band tunnelling probabilities are calculated based on the Wentzel-Kramers-Brillouin approximation using a two-band dispersion relation.
Finally, ohmic contacts on the top and the bottom of the device are considered with the large carrier extraction velocity of $v_\mathrm{cont} = \SI[per-mode=symbol]{1e7}{\centi\metre\per\second}$ on the nanowire-contact interface for both majority and minority carriers.


\subsection{Nanowire junction optimization}
\label{sec:ElRes}

\begin{figure*}[ht]
\centerline{
\includegraphics[width=0.4\columnwidth]{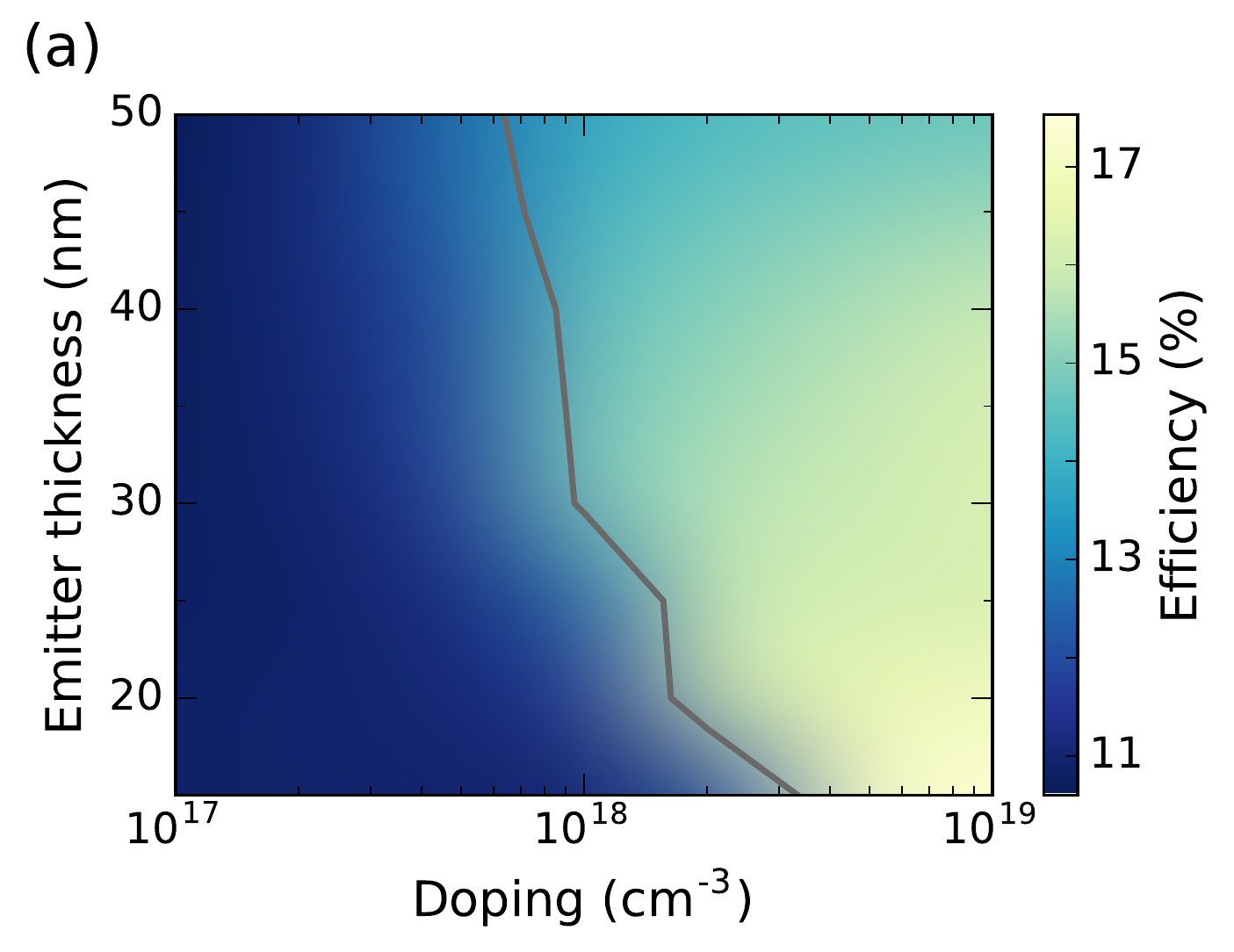} \hspace{10mm}
\includegraphics[width=0.4\columnwidth]{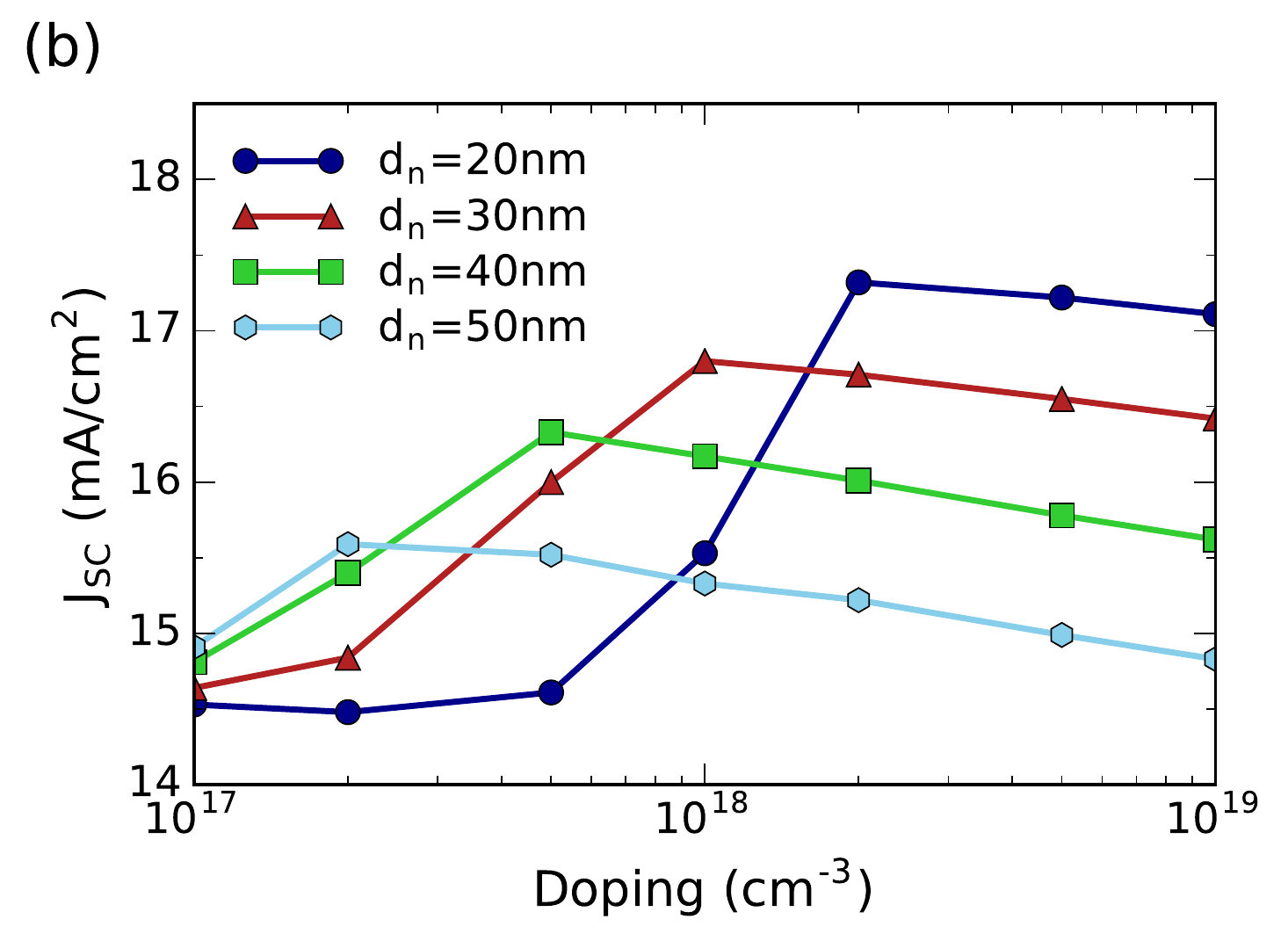}
}
\centerline{
\includegraphics[width=0.4\columnwidth]{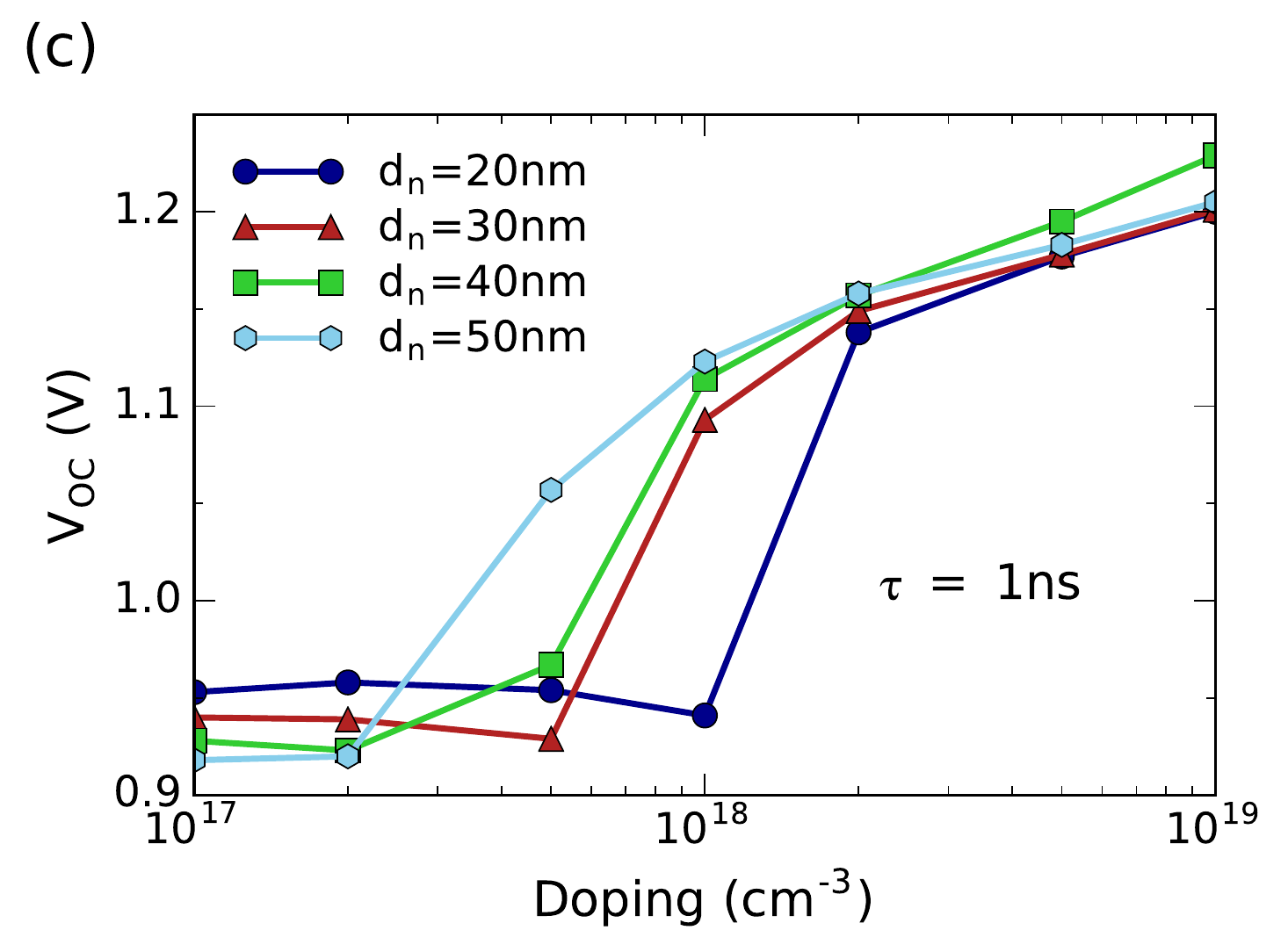} \hspace{10mm}
\includegraphics[width=0.4\columnwidth]{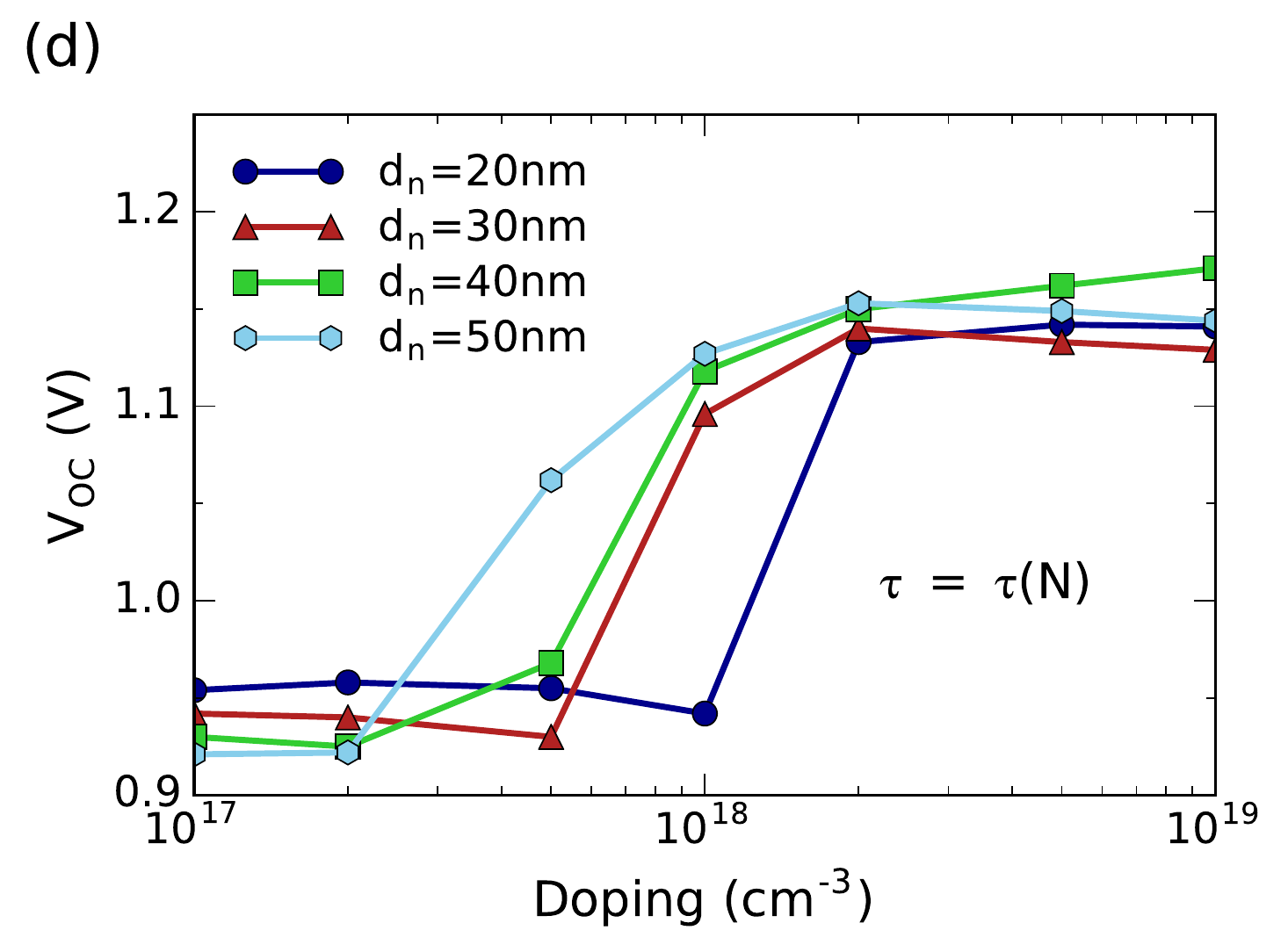}
}
\caption{Results of electrical simulations of the top cell. (a) Power conversion efficiency color map for different doping concentration and shell thickness of the junction. Solid line shows the boundary position of the completely depleted shell regime. Panels (b) and (c) show variation of the key elements: $J_\mathrm{SC}$ and $V_\mathrm{OC}$ with $ p - n$ junction parameters for the case of constant SRH lifetime $\tau$ = \SI{1}{\nano\second}. (d) Variation of $V_\mathrm{OC}$ for doping-dependent minority carrier lifetime according to Eq.~(\ref{GaAs_lifetime_1}) - (\ref{GaAs_lifetime_2}). }
\label{fig:ElOpt_const}
\end{figure*}

As electrical properties of the core-shell nanowire cell are substantially more complex than the ones of the silicon planar solar cell, the electrical modelling of the top subcell is first carried out. To study charge transfer through the nanowires only, an auxiliary contact is added at the bottom of the nanowire (represented by the dashed line in Fig.~\ref{fig:el_system}). 
Radial $p-n$ junction in principle allows for a more efficient carrier collection \cite{Yao14} and is less prone to surface recombination \cite{Huang12} than the axial one. The downside is that its geometry and doping level must be chosen carefully in order to avoid depletion of either core or shell \cite{Chia13, Li15b}. The design process is additionally complicated by the fact that the quality of the surface passivation also impacts the choice of optimal parameters, as charge accumulation on the surface enhances shell depletion \cite{Chia12}. In this section the results of the electrical optimization of the top junction are presented.

Emitter thickness $d_n$, which is taken to be the same on the top and on the side shell of the NW, is varied together with the doping of the junction. In this study doping concentration in the core and the shell are taken equal, so that $N_p = N_n = N$. 
SRH minority carrier lifetime and SRV are kept constant, independent on the doping level and equal to $\tau = \SI{1}{\nano\second}$ and $S = \SI[per-mode=symbol]{1e4}{\centi\metre\per\second}$ (see table \ref{tab:el} for more details).
Power conversion efficiency of the sunlight illuminating the nanowire array $\eta_\mathrm{NW} = P_\mathrm{NW}/P_\mathrm{inc}$ serves as the main figure of merit. However, it should be noted that for tandem solar cell application, high $J_\mathrm{SC}$ should be prioritized due to the current matching condition of the series cell connection. 

The results are summarized in Fig.~\ref{fig:ElOpt_const} (a-c), which plots NW subcell efficiency $\eta$, short-circuit current $J_\mathrm{SC}$ and open-circuit voltage $V_\mathrm{OC}$ as functions of doping and shell thickness. 
A considerable increase of efficiency is achieved through the increase of both $J_\mathrm{SC}$ and $V_\mathrm{OC}$ at high $N$. 
It is explained as follows: charge accumulation and band bending on the NW surface may lead to the complete depletion of the thin shell for shallow junctions, resulting in the severe reduction of $V_\mathrm{OC}$ and increase of SRH recombination \cite{Chia13}. 
For even lower doping concentrations, an intrinsic shell with an effectively axial $p-n$ junction can be observed, with the exact depth of the junction determined by the top contact geometry.

\begin{figure*}[t]
\centerline{
\includegraphics[width=0.18\columnwidth]{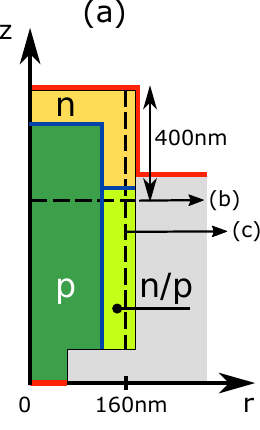} \hspace{5mm}
\includegraphics[width=0.36\columnwidth]{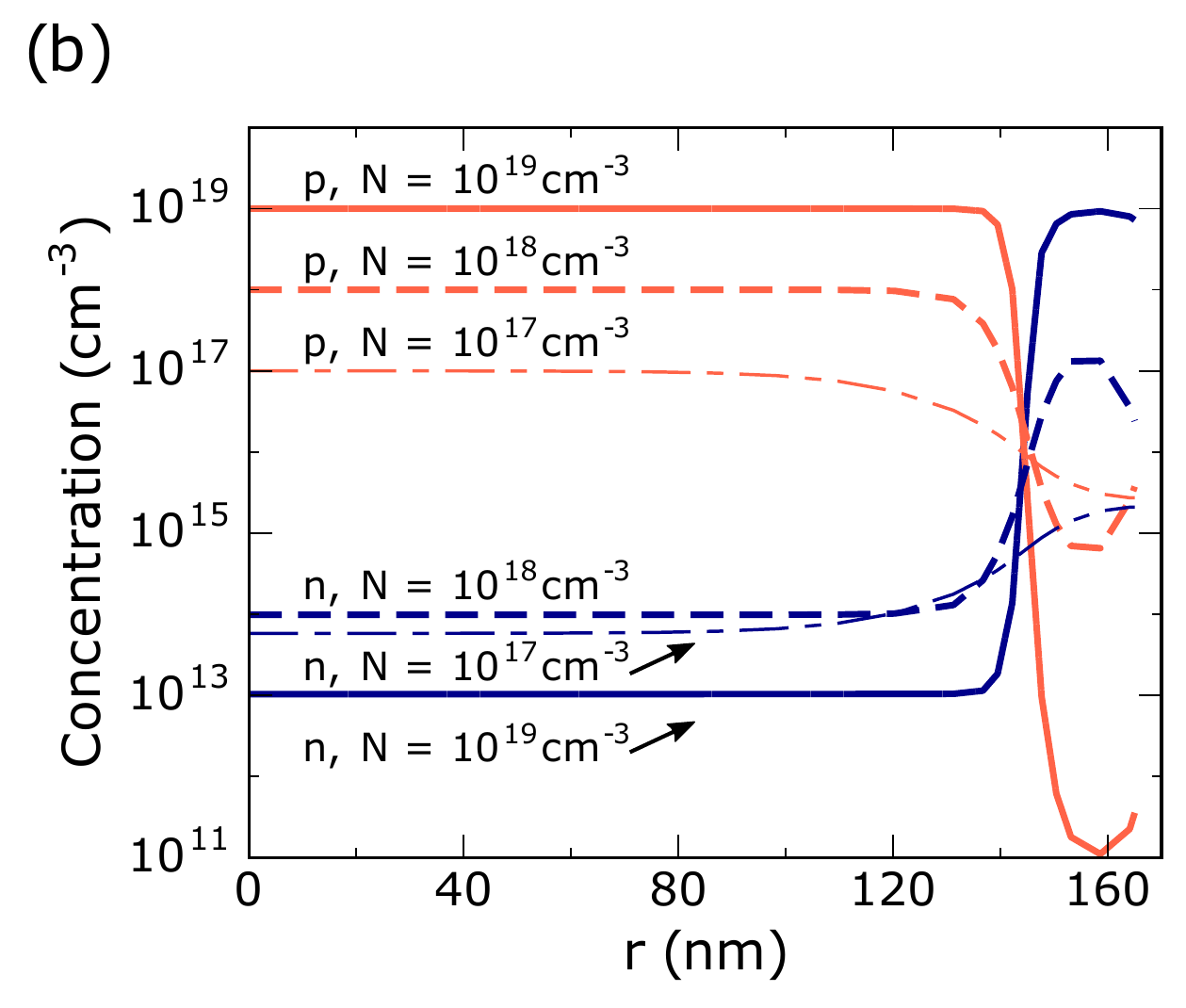} \hspace{5mm}
\includegraphics[width=0.36\columnwidth]{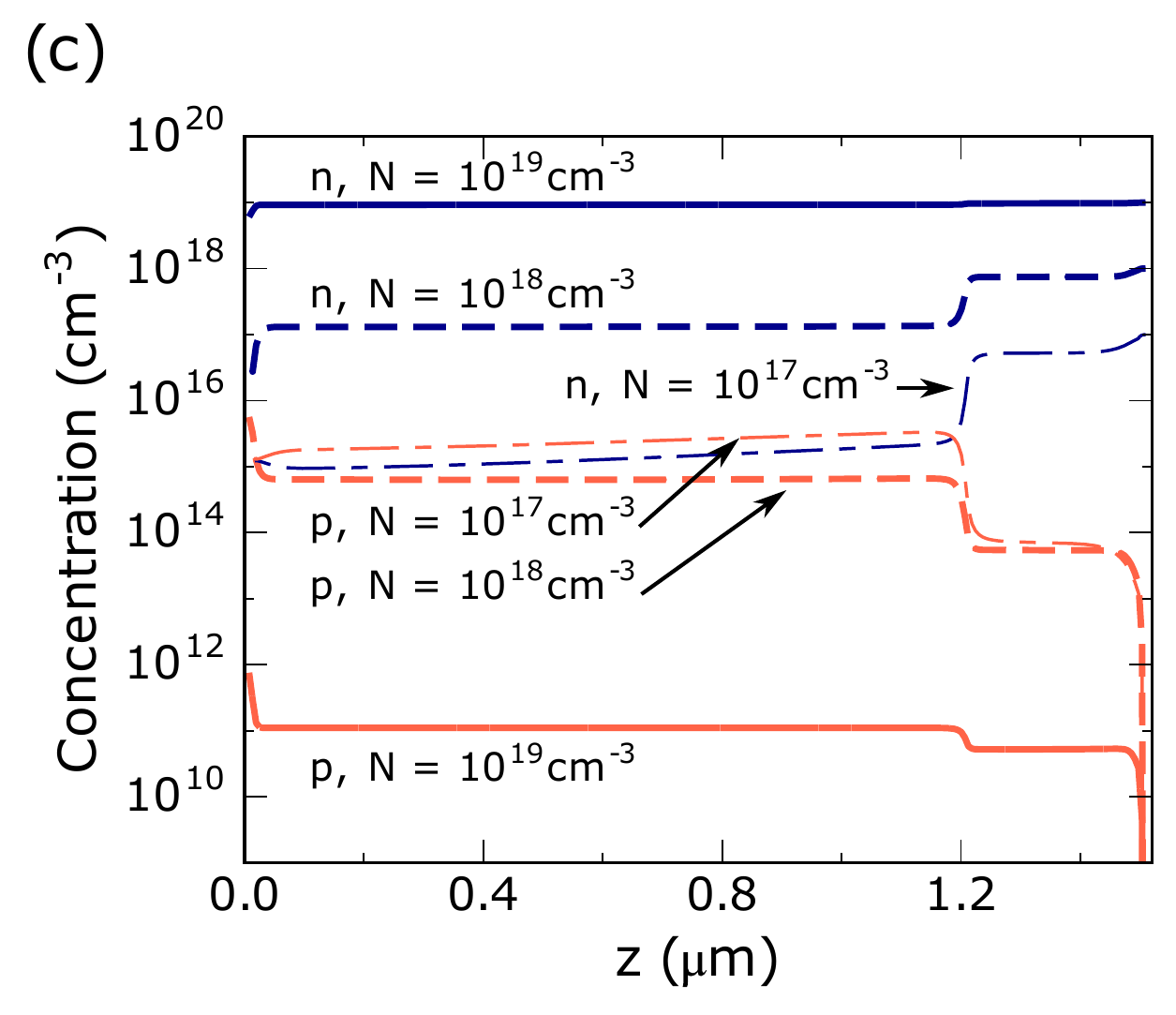} \hspace{5mm}
}
\caption{(a) Scheme of transition from radial $p-n$ junction to effectively axial one, when the carrier inversion occurs in the bottom part of the nanowire shell. 
(b) Carrier concentration profiles for the horizontal cut \SI{0.4}{\micro\metre} deep from the top of the NW. Three pairs of lines correspond to electron (blue) and hole (red) densities for different doping of the radial junction.
(c) Same for the vertical cut at $r = \SI{160}{\nano\metre}$. Both cuts are marked by the dashed lines on the scheme (a).}
\label{fig:density_cuts}
\end{figure*}

The transition from the radial to axial junction regime is schematically illustrated in Fig.~\ref{fig:density_cuts} (a). Solid blue lines indicate the junction positions for the two limiting cases of high and low doping. 
On panels (b) and (c) the calculated carrier concentration profiles are plotted at constant $h$ and $z$ for the $d_n = \SI{20}{\nano\metre}$ thick shell.
Three pairs of lines on each panel correspond to the electron (blue lines) and hole (red lines) densities for three doping concentrations $N = \SI{1e17}{\centi\metre^{-3}}$, \SI{1e18}{\centi\metre^{-3}} and \SI{1e19}{\centi\metre^{-3}}.
For the intermediate case, the shell is already fully depleted, and for the lowest doping, a carrier inversion phenomenon is even observed.
This underlines the necessity of growing a thicker or more heavily doped junctions. Decreasing the core doping is another solution \cite{Chia13}, which has a risk, however, of the complete depletion of the core, especially for the nanowires with smaller radius.

The critical doping values $N_\mathrm{crit}$, below which the shell is completely depleted, are extracted from the drop of the open-circuit voltage (Fig.~\ref{fig:ElOpt_const} (c)) and plotted in Fig.~\ref{fig:ElOpt_const} (a) by the solid line.
In the rest of the section, parameter region above $N_\mathrm{crit}$ (alternatively, $d_\mathrm{crit}$) is studied and more precise electrical optimization is performed.
First, higher $J_\mathrm{SC}$ is observed for thinner emitters (Fig.~\ref{fig:ElOpt_const} (b)). Indeed, thicker shells exhibit decreased carrier collection properties as more holes may recombine at the top contact or at the surface of the nanowire. In addition to this, the increase of doping reduces the depletion width of the junction, which leads to slightly worse minority carrier extraction from the emitter.
Despite this, efficiency $\eta$ continues to grow with the doping of the NW due to the reduction of the highly recombining depletion region, as can be seen from the increase of $V_\mathrm{OC}$ with doping in Fig.~\ref{fig:ElOpt_const} (c)). 

The situation changes, however, if degradation of the carrier lifetime with doping is taken into account. There is yet no general agreement on the form or even the existence of doping dependence of $\tau$ in AlGaAs \cite{Neumann93}, so our predictions are based on the corresponding relation for pure bulk GaAs:
\begin{align}
\tau_n &= \tau_0\left(\frac{\SI{1.3e9}{\centi\metre^{-3}}}{N_p}\right)^{0.90}, \label{GaAs_lifetime_1} \\ 
\tau_p &= \tau_0\left(\frac{\SI{4.0e9}{\centi\metre^{-3}}}{N_n}\right)^{0.92}.
\label{GaAs_lifetime_2}
\end{align}
This formula with the prefactor $\tau_0 = \SI{1}{\second}$ gives a good fit to the available data for GaAs in the doping interval $\SI{1e16}{\centi\metre^{-3}} \leq N \leq \SI{1e19}{\centi\metre^{-3}}$ relevant for the current study, \cite{Adachi05}. For AlGaAs nanowires, we use $\tau_0 = \SI{0.1}{\second}$ to account for possible lifetime deterioration in the nanowires as well as due to the AlAs fraction in the alloy.
Equations (\ref{GaAs_lifetime_1}) - (\ref{GaAs_lifetime_2}) then give a reasonable $\tau_n = \SI{1}{\nano\second}$ and $\tau_p = \SI{1.87}{\nano\second}$ at $N = \SI{1e18}{\centi\metre^{-3}}$.
Fig.~\ref{fig:ElOpt_const} (d) shows the behavior of the $V_\mathrm{OC}$ for the case of doping-dependent lifetime. 
Voltage saturation is observed at high $N$ in contrast with the monotonic increase of $V_\mathrm{OC}$ in the case of constant $\tau$.
As a result, efficiency $\eta$ now features a broad maximum at average doping level $N \simeq \SI{3e18}{\centi\metre^{-3}}$ (not shown here). 
Therefore, in the rest of the paper $d_n = \SI{20}{\nano\metre}$ and $N_n = N_p = \SI{5e18}{\centi\metre^{-3}}$ are used for the optimized top junction parameters. The impact of the two main recombination processes on the top cell performance is studied in the next section.


\subsection{SRH recombination}
\label{ssec:SRH}

Shockley-Read-Hall recombination lifetime varies greatly in different AlGaAs devices \cite{Neumann93}.
Moreover, as $\tau$ is related to the concentration of defects in the structure, it also depends strongly on the parameters of the growth process. Therefore, it is important to understand the influence of $\tau$ on the overall cell efficiency in the wide range of SRH lifetimes. 
For this study optimal junction parameters from section \ref{sec:ElRes} $d_n = \SI{20}{\nano\metre}$ and $N = \SI{5e18}{\per\centi\metre\cubed}$ are used.

\begin{figure}[ht]
\centering
\includegraphics[width=0.4\columnwidth]{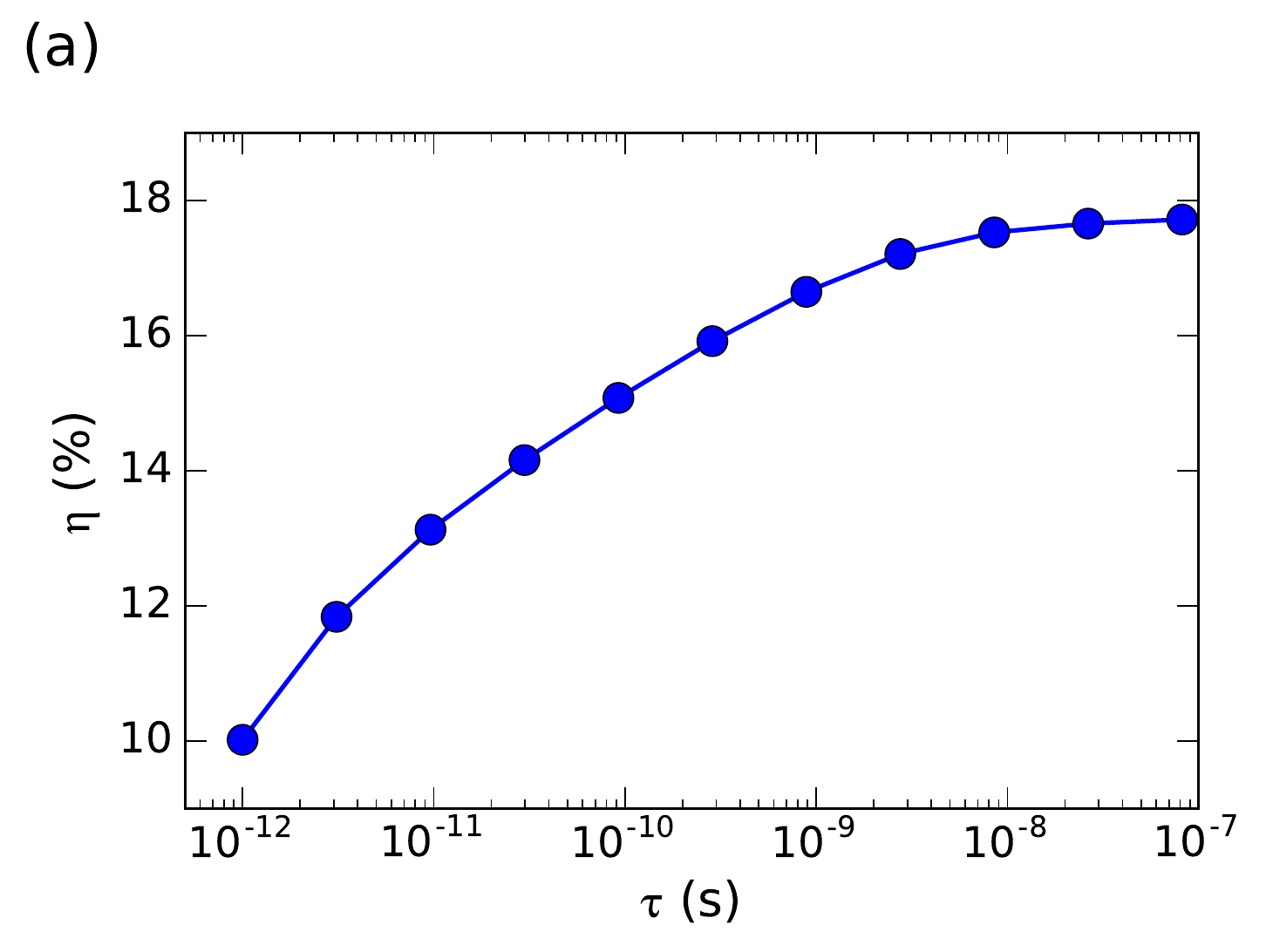} \vspace{2mm}
\includegraphics[width=0.45\columnwidth]{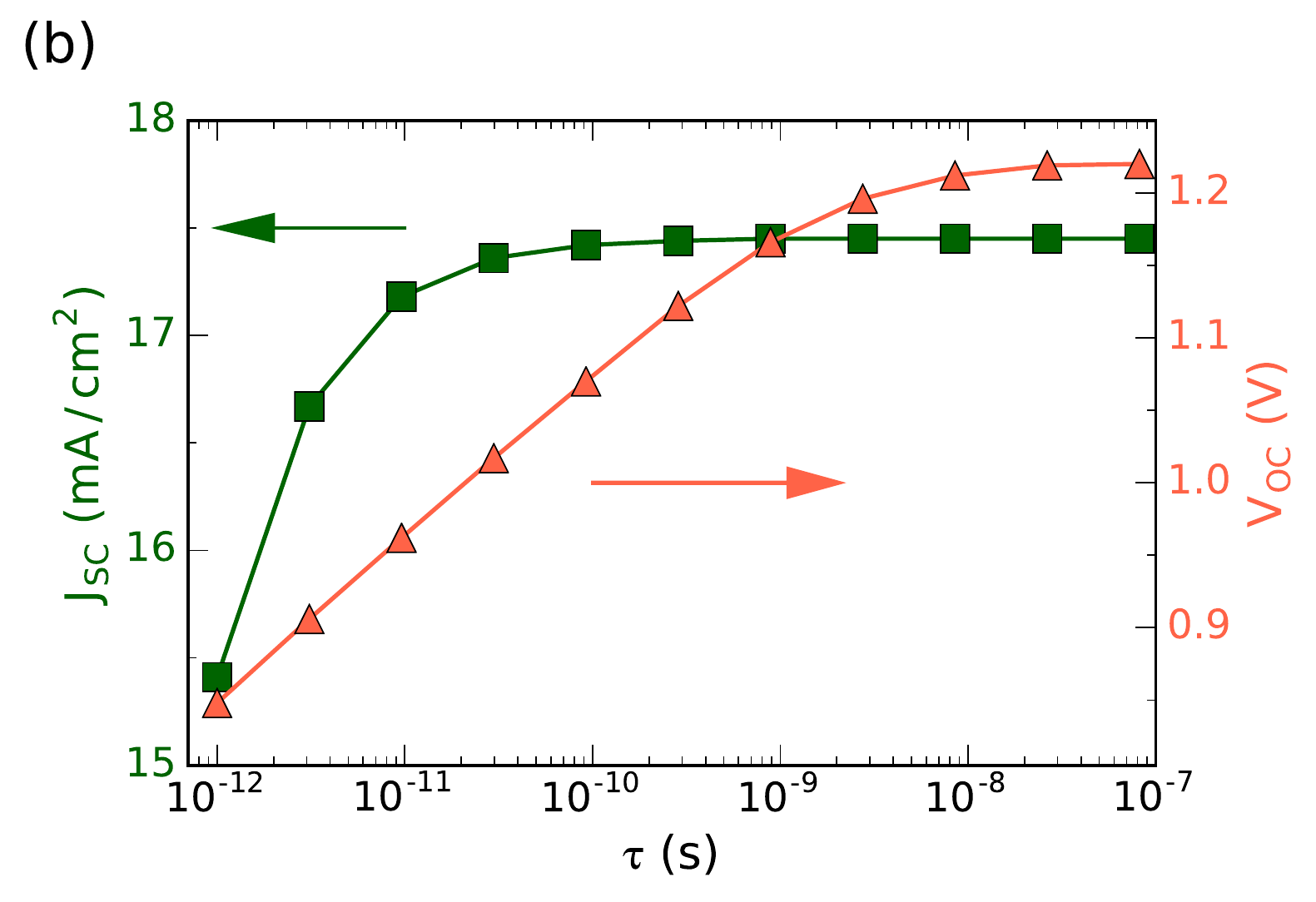}
\caption{Performance of the top cell ($d_n = \SI{20}{\nano\metre}$, $N = \SI{5e18}{\per\centi\metre\cubed}$) with varying SRH recombination lifetime.}
\label{fig:SRH}
\end{figure}

The variation of $\tau$ within several decades shows logarithmic increase of cell efficiency (Fig.~\ref{fig:SRH}). At very high $\tau \geq \SI{10}{\nano\second}$, efficiency saturates reflecting that recombination on defects gives ground to radiative recombination as the dominant effect. An important result is that short circuit current is not affected by the SRH recombination for $\tau > \SI{0.1}{\nano\second}$. Constant $J_\mathrm{SC}$ in the wide range of realistic minority carrier lifetimes means that bulk defects do not influence the current matching condition, which is a very favorable result for tandem solar cells. 
Similar behavior was observed by Wang et al. \cite{Wang15} in their study of GaInP-NW/Si tandem structure, which evidences that excellent carrier collection characteristics of the system originate from the general properties of the radial junction in the NW rather than to specific material or design.


\subsection{Surface recombination}
\label{ssec:SRV}

\begin{figure}[ht]
\centering
\includegraphics[width=0.4\columnwidth]{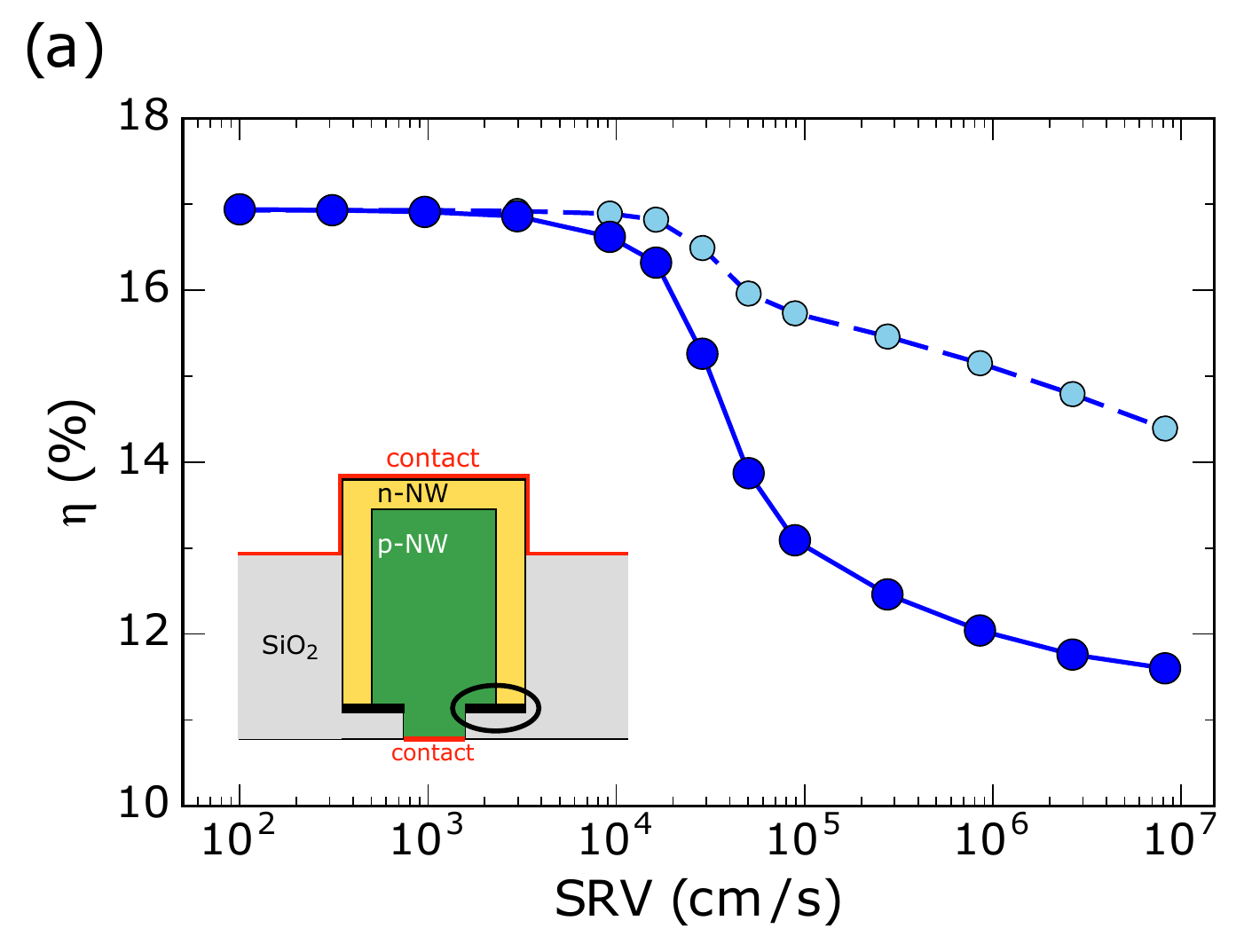} \vspace{5mm}
\includegraphics[width=0.45\columnwidth]{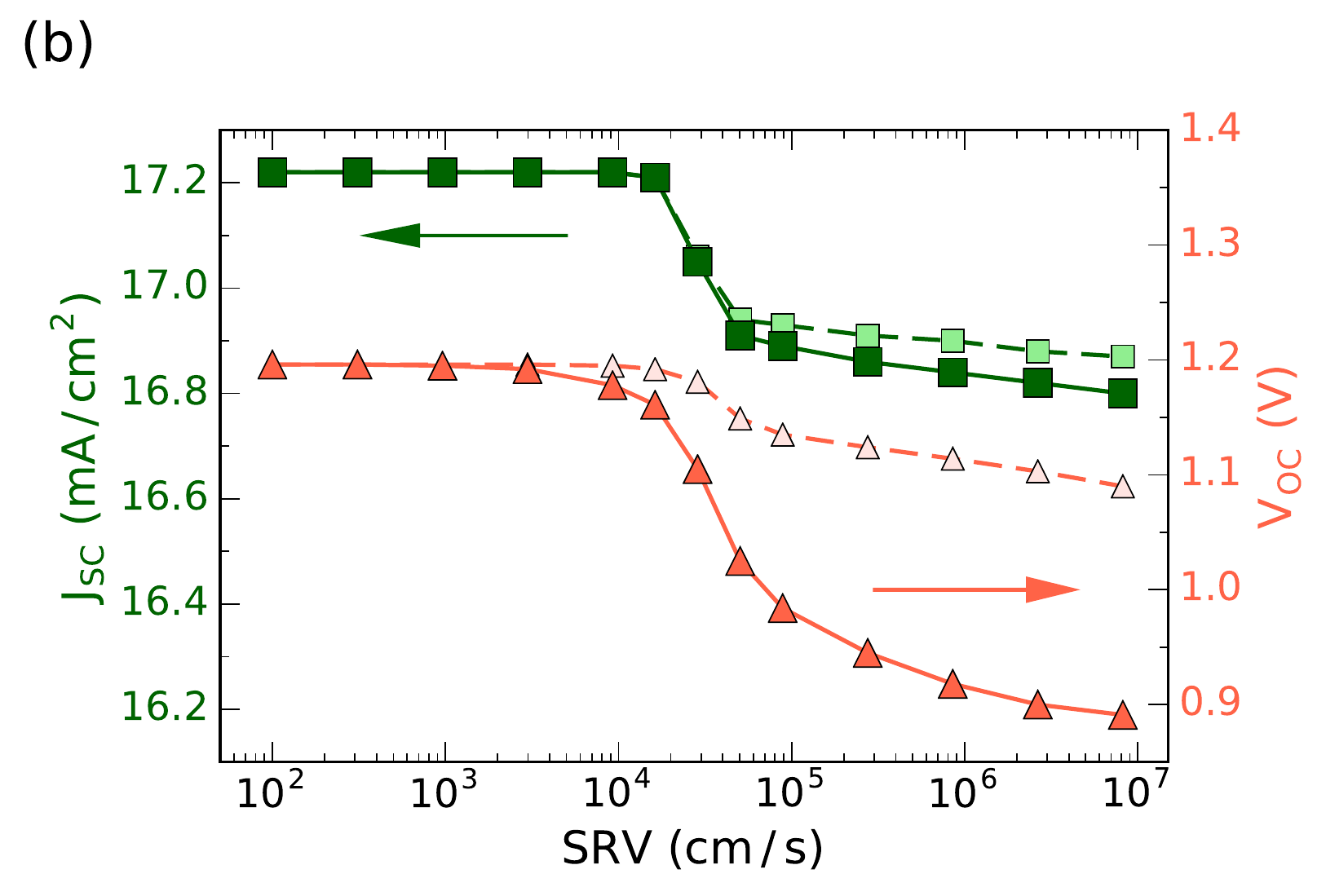}
\caption{Dependence of the top cell ($d_n = \SI{20}{\nano\metre}$, $N = \SI{5e18}{\per\centi\metre\cubed}$) performance on the simultaneous variation of surface recombination velocity and surface trap density. Discontinuous lines show the results of the simulations of the perfectly passivated bottom AlGaAs - $\mathrm{SiO_2}$ interface (shown by thick black line on the inset). 
}
\label{fig:SRV}
\end{figure}

As mentioned before, the value of SRV strongly affects the optimal parameters of radial $p-n$ junction \cite{Chia13}. So for the needs of electrical modelling $S$ should also be regarded as an important independent parameter. 
In section \ref{sec:ElRes}, electrical optimization for fixed $S$ was performed. The effect of varying surface trap density and recombination velocity on the optimal configuration is studied in the following.
SRV is varied within $ \SI[per-mode=symbol]{1e2}{\centi\metre\per\second} \leq S \leq \SI[per-mode=symbol]{1e7}{\centi\metre\per\second}$, the range of observed recombination rates for different $\mathrm{Al_xGa_{1-x}As}$ interfaces \cite{Pavesi94, Jiang12}.
It is important to remind that trap cross section $\sigma_s$ and thermal velocity $v_\mathrm{th}$ are fixed throughout this work, so surface defect density is varied together with $S$ according to Eq.~(\ref{SRV}).
The results are plotted in Fig.~\ref{fig:SRV} with solid lines. 

The behavior of the cell performance with different $S$ can be well explained by shell depletion and carrier inversion mechanisms, as has been presented in section \ref{sec:ElRes}.
At low $S\leq\SI[per-mode=symbol]{2e4}{\centi\metre\per\second}$ surface recombination adds only a small contribution to the bulk and contact recombination processes. 
A severe decrease in cell performance happens at $S = \SI[per-mode=symbol]{5e4}{\centi\metre\per\second}$ due to to the appearance of a highly recombining quasi-intrinsic layer and indicates the depletion of the NW shell.

In addition to the general analysis of surface recombination in the nanowire subcell, the contributions of different parts of AlGaAs-insulator interface are studied separately by artificially turning off recombination at the base of the NW $S_\mathrm{bot} = 0$ (see black line in the inset of Fig.~\ref{fig:SRV} (a)). The motivation for this is the fact that outer surface of the NW is often treated by various passivation layers to reduce its SRV. The bottom part, however, is an interface with the insulating mask and is processed differently during the NW growth. 
The results for the case of perfectly passivated bottom interface are shown in Fig.~\ref{fig:SRV} by dashed lines. The difference between the corresponding pair of lines is then attributed to recombination losses on the bottom NW surface.
It is clear that despite its very small size (bottom part constitutes around $5\%$ of the total NW surface in our geometry), the base of the NW is as important as the rest of the surface. 
In the core-shell junction design, bottom facet is the region where the depletion zone is in contact with the NW surface.
Therefore the fact that a greater part of $V_\mathrm{OC}$ loss occurs at the interface with the growth mask is in agreement with the works reporting enhancement of surface recombination, localized in the depletion region \cite{LaPierre11a, Kuhn00, Kessler12, Bertrand17}.

The results of this subsection are summarized as follows. 
The performance of the insufficiently passivated nanowires is severely decreased with the main contribution coming from the $V_\mathrm{OC}$ degradation similarly to the SRH recombination on the bulk defects. 
A large part of this degradation occurs in a spatially small part of the surface, which emphasizes the importance of careful treatment of the bottom part of the radial $p-n$ junction in both modelling and experiment.
However, for the passivated NW surface with $S \leq \SI[per-mode=symbol]{1e4}{\centi\metre\per\second}$ and $N_\mathrm{TD} = N_\mathrm{TA} = \SI[per-mode=symbol]{1e12}{\per\centi\metre\squared}$ surface recombination does not lead to severe efficiency decrease. 
Such high quality nanowire surfaces were already achieved for passivated GaAs nanowires \cite{Songmuang16, Jiang12}, and should be used in high efficiency nanowire solar cells.
Finally, it was shown that surface defects do not directly severely alter the current-matching condition as long as the $p-n$  junction shell stays in the proper partially depleted regime.


\subsection{Tandem cell}
To evaluate the cumulative impact of all optimizations on the overall efficiency, the simulations on the full tandem structure are performed. 
Bottom subcell consists of a highly doped \SI{15}{\nano\metre} thin $n^{++}$ emitter  and a \SI{200}{\micro\metre} $p$-doped monocrystalline silicon collector.
Another highly doped \SI{15}{\nano\metre} $p^{++}$ layer on top of the subcell forms a tunnel junction, which provides transparent ohmic contact with the top cell.
The back surface field layer in Si, is modelled by setting electron extraction velocity on the bottom contact to zero. 
Other simulation parameters are gathered in the table \ref{tab:el}. 

\begin{figure}[ht]
\centerline{
\includegraphics[width=0.45\columnwidth]{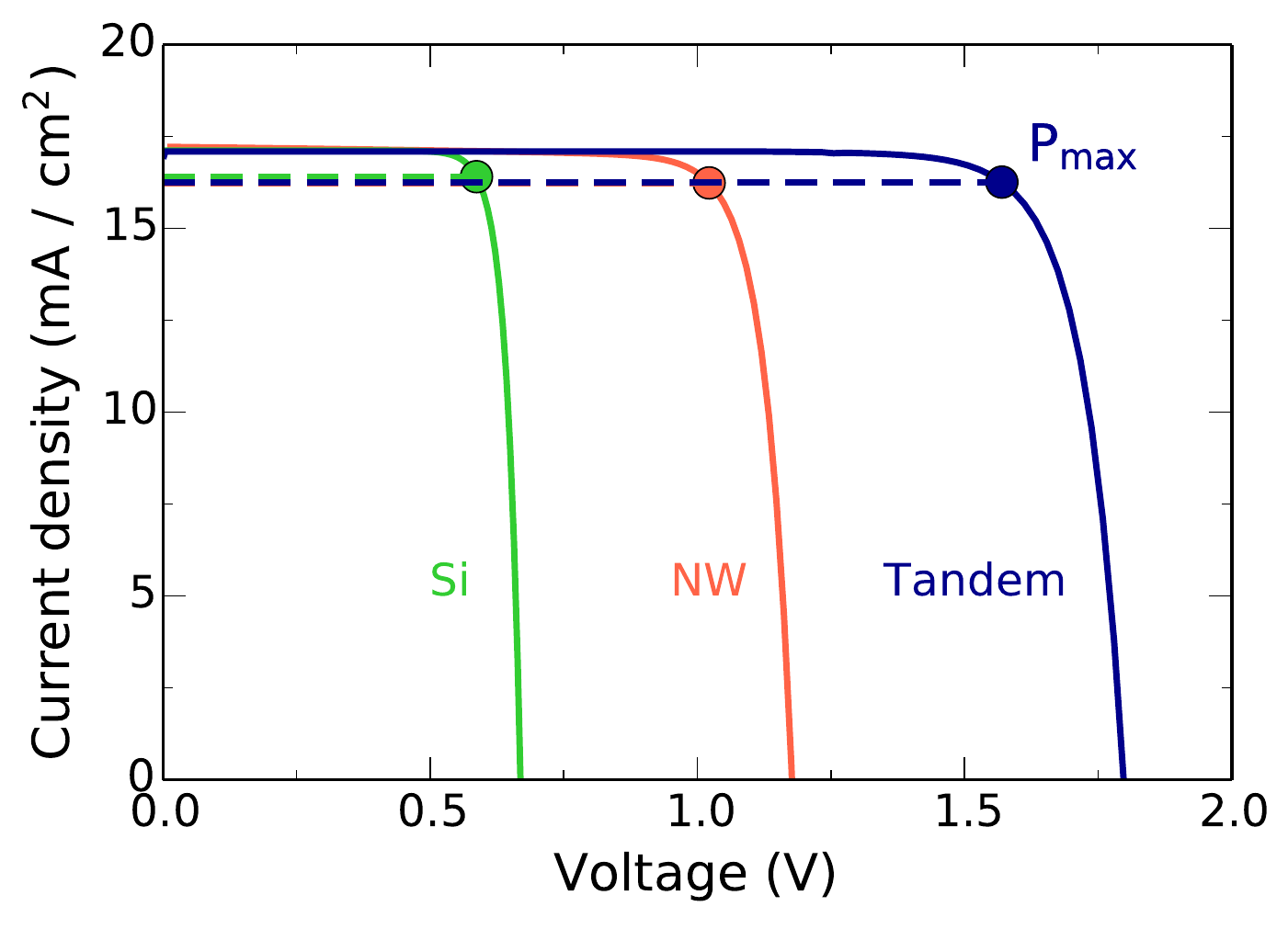}
}
\caption{$J-V$ characteristics for the final optimized tandem solar cell with efficiency $\eta = 27.6\% $. Maximum power output points and corresponding working current are indicated.}
\label{fig:J-V_tandem}
\end{figure}


Figure \ref{fig:J-V_tandem} shows the $J-V$ characteristics of the two separate subcells, together with the resulting curve for the full tandem structure. 
First, it can be noted that the current at the maximum power point of the tandem cell is only \SI[per-mode=symbol]{0.2}{\milli\ampere\per\centi\metre\squared} below the one of the Si subcell.
Figure \ref{fig:J-V_tandem} also shows that the resulting $V_\mathrm{OC}$ is almost the sum of the open circuit voltages of the subcells simulated alone. The difference of \SI{50}{\milli\volt} is attributed to the tunnel junction and to the slightly lower current in top subcell. 

Finally, no variation of tandem cell performance with the mask opening diameter is observed within the range $\SI{50}{\nano\metre} \geq d_\mathrm{mask} \geq \SI{200}{\nano\metre}$.
The studied AlGaAs-NW/Si tandem cell is able to reach a promising efficiency of $\eta = 27.6\%$ with $J_\mathrm{SC} = \SI[per-mode=symbol]{17.1}{\milli\ampere\per\centi\metre\squared}$,  $V_\mathrm{OC} = \SI{1.85}{\volt}$ and fill factor $\mathrm{FF} = 0.87$.


\section{Conclusion}

Coupled optoelectronic simulations of the dual-junction $\mathrm{Al}_{0.2}\mathrm{Ga}_{0.8}\mathrm{As}$ nanowire on silicon solar cell have been performed. 
Highly efficient numerical methods allow modelling a  tandem structure with various auxiliary layers taken into account: ITO, passivation, dielectric growth mask, tunnel diode. 
The studied system is promising for achieving high power conversion efficiencies, the inherent complexity, however, reveals itself in the sensitivity to the variation of its numerous parameters. 
Providing the following set of cell design rules, this work could serve as a guideline for growth and experimental effort.

First, optical simulations show that the height of the III-V nanowires with the close-to-optimal bandgap of $E_g = \SI{1.7}{\electronvolt}$ can be restricted to $H = \SI{1.5}{\micro\metre}$. Further growth only slightly improves matched photogenerated current.
Then, a wide range of array geometries with $D/p \geq 0.5$ are shown to have acceptable total photocurrent, only $5\%$ lower than the optimal one.
But exact array configuration is sensitive to the geometry and materials used. In order to achieve maximal photocurrent in the tandem cell, precise optimization is necessary for every particular case.
Finally, optical simulations reveal the effect of ITO as an antireflection layer for nanowire arrays with large III-V filling ratios.

Core-shell design of the top junction can be beneficial for the top cell, but to achieve peak performance, it also requires a precise optimization. 
Electrical simulations demonstrate the paramount importance of doping the junction enough or alternatively making sufficiently thick core and shell of the nanowire to avoid their full depletion.
A more refined optimization shows that, once above the critical doping level, thin moderately doped shells provide best performance of radial junction in the nanowire solar cell. 
Surface passivation also plays an important role as surface charge accumulation leads to the shell depletion.
But once in a proper operating regime, trap-assisted recombination  on the surface has limited effect on the photocurrent density. The same is true for the bulk SRH recombination in the nanowire. These properties of the radial $p-n$ junction are very beneficial for tandem cell application. 
To reach maximum efficiency high but reasonable carrier lifetime of $\tau \simeq \SI{1}{\nano\second}$ and surface recombination velocity $S \simeq \SI[per-mode=symbol]{1e4}{\centi\metre\per\second}$ are required.
Another important conclusion is that in the core-shell geometry a special attention should be put to the passivation of the base of the nanowire as recombination processes on this small interface with the growth mask are enhanced due to the contact with the depleted region of the top junction. 

The necessity to precisely control a large number of important parameters presents a substantial technological challenge. 
Nevertheless, the results of this study demonstrate that following the presented guidelines, makes possible the construction of multi-junction solar cells based on III-V nanowires with efficiency over 27\%.


\appendix

\section{Generation rate calculation in silicon}
\label{sec:App}
\begin{figure}[ht]
\centerline{
\includegraphics[width=0.45\columnwidth]{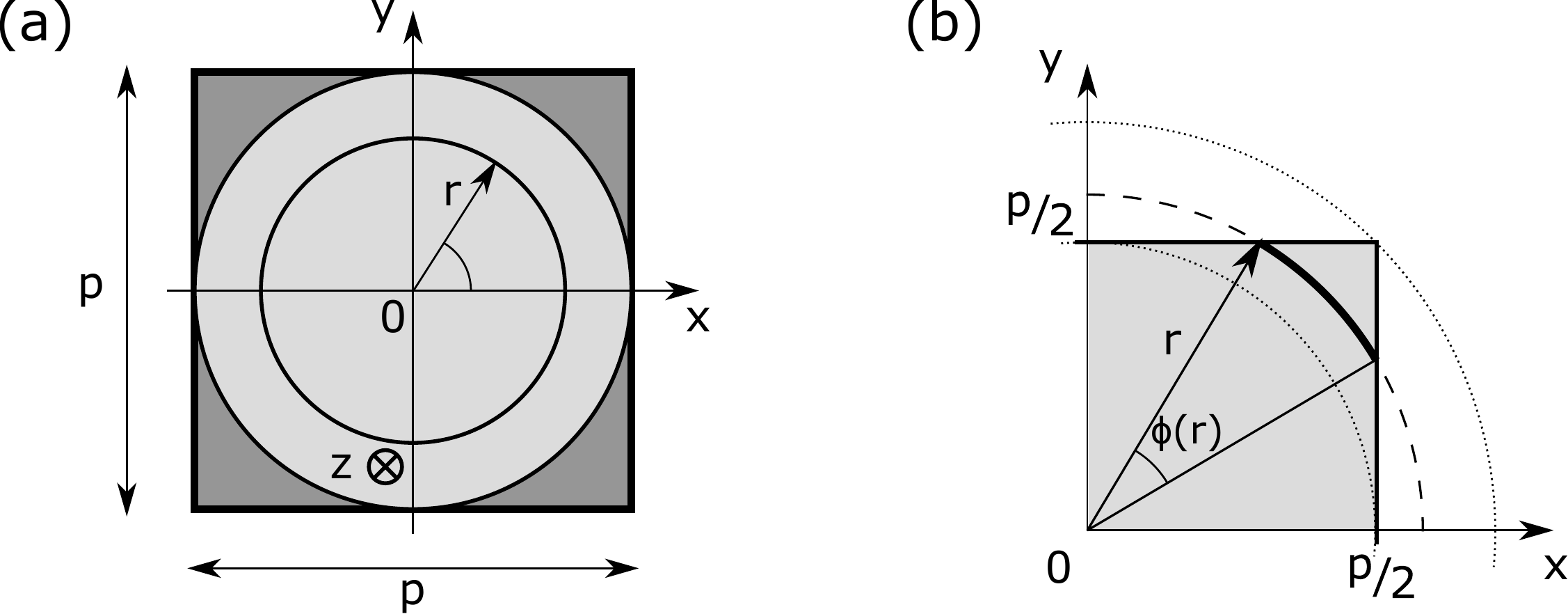}
}
\caption{(a) Schematic of the horizontal cut in the silicon substrate with a square unit cell and planar coordinate system.  (b) The generalization of the angle-averaging procedure for $r > p/2$ used to calculate planar generation rate $G(r, z)$ in the silicon substrate.}
\label{fig:renorm_scheme}
\end{figure}

Carrier generation rate in a single nanowire, which serves to couple optical and electrical simulations, is usually calculated using equations (\ref{GR}) and (\ref{angle-averaging}). 
The procedure is described in detail by  Michallon \emph{et al.} \cite{Michallon14}, it uses the approximation of a square symmetry of the NW array by a cylindrical symmetry of a single nanowire. 
Nevertheless, the angle averaging (\ref{angle-averaging}) is justified, as in such wave guide variation of electric field as a function of polar angle is minimal compared to radial and vertical directions.
This appendix describes the generalization of the polar angle averaging for the case of a cubic unit cell   
since direct application of Eq.~(\ref{angle-averaging}) leads to the underestimation of the photogenerated current. Figure \ref{fig:renorm_scheme} (a) shows a horizontal cut of the Si substrate at any fixed quota $z=z_0$ with the in-plane coordinates $-p/2\leq x \leq p/2$ and $-p/2\leq y \leq p/2$. It illustrates that the corners of a square unit cell are not taken into account if a usual angle-averaging scheme is used.

First, the averaging procedure is modified by employing an enlarged radial coordinate $r \in [0, p/\sqrt{2}]$. Then, to avoid double counting on the sides of the square, averaging of the electric field is done only for the points inside the square unit cell of the substrate: 
\begin{equation}
 \vert \mathbf{E}(r,z)\vert^2 = \frac{\phi(r)}{4\pi^2}\int\displaylimits_\mathrm{u.\,c.}\vert \mathbf{E}(r, \theta, z)\vert^2d\theta.
\end{equation}
Here $\phi(r)$ has a meaning of an inner angle of the arc with the radius $r$, which falls inside the original square unit cell (thick line in Fig.~\ref{fig:renorm_scheme} (b)). This renormalization function can be calculated as
\begin{equation}  
\phi(r) = 
     \begin{cases}
      2\pi; &\ r < \frac{p}{2}, \\    
      4\arccos\Big(\frac{p}{r}\sqrt{1-\frac{p^2}{4r^2}}\Big); &\  \frac{p}{2} \leq r \leq \frac{p}{\sqrt{2}}, \\
			0; &\ \frac{p}{\sqrt{2}} < r.            
     \end{cases}
\label{SiAveraging}
\end{equation}
The generation rate determined in such a way conserves total photogenerated current in the substrate, similarly to the averaging procedure in the nanowire (\ref{angle-averaging}). 
It was employed to obtain the dual-dimensional distribution of the generation rate in silicon (Fig.~\ref{fig:GenRate} (b)), which is further used in the tandem cell electrical simulations.
A visible sharp peak of $G(r, z)$ at $r=p/2=\SI{275}{\nano\metre}$ is precisely an artefact of the renormalization procedure, described above.
It illustrates the fact, that the assumption of the weak variation of generation rate in the polar direction is not correct for $r\gtrsim p/2$, where the true square symmetry of the system is most pronounced.
Nevertheless, the use of Eq.~(\ref{SiAveraging}) gives a suitable spatial distribution of the generation rate, especially for $r<p/2$, where the majority of the absorption takes place.


\section*{Acknowledgement}
The authors thank Alain Fave and Michel Gendry for fruitful discussions.
This work was supported by the French National Research Agency within the framework of the HETONAN project (ANR-15-CE05-0009-04).


\section*{References}
\bibliography{Tandem}

\begin{thebibliography}{10}
\expandafter\ifx\csname url\endcsname\relax
  \def\url#1{\texttt{#1}}\fi
\expandafter\ifx\csname urlprefix\endcsname\relax\def\urlprefix{URL }\fi
\expandafter\ifx\csname href\endcsname\relax
  \def\href#1#2{#2} \def\path#1{#1}\fi

\bibitem{Garnett11}
E.~C. Garnett, M.~L. Brongersma, Y.~Cui, M.~D. McGehee, Nanowire {Solar}
  {Cells}, Annual Review of Materials Research 41~(1) (2011) 269--295.
\newblock \href {http://dx.doi.org/10.1146/annurev-matsci-062910-100434}
  {\path{doi:10.1146/annurev-matsci-062910-100434}}.

\bibitem{Kupec10}
J.~Kupec, R.~L. Stoop, B.~Witzigmann, Light absorption and emission in nanowire
  array solar cells, Opt. Express, OE 18~(26) (2010) 27589--27605.
\newblock \href {http://dx.doi.org/10.1364/OE.18.027589}
  {\path{doi:10.1364/OE.18.027589}}.

\bibitem{Wallentin13}
J.~Wallentin, N.~Anttu, D.~Asoli, M.~Huffman, I.~\r{A}berg, M.~H. Magnusson,
  G.~Siefer, P.~Fuss-Kailuweit, F.~Dimroth, B.~Witzigmann, H.~Q. Xu,
  L.~Samuelson, K.~Deppert, M.~T. Borgstr\"{o}m, {InP} {Nanowire} {Array}
  {Solar} {Cells} {Achieving} 13.8\% {Efficiency} by {Exceeding} the {Ray}
  {Optics} {Limit}, Science 339~(6123) (2013) 1057--1060.
\newblock \href {http://dx.doi.org/10.1126/science.1230969}
  {\path{doi:10.1126/science.1230969}}.

\bibitem{Green17}
M.~A. Green, K.~Emery, Y.~Hishikawa, W.~Warta, E.~D. Dunlop, D.~H. Levi,
  A.~W.~Y. Ho-Baillie, Solar cell efficiency tables (version 49), Prog.
  Photovolt: Res. Appl. 25~(1) (2017) 3--13.
\newblock \href {http://dx.doi.org/10.1002/pip.2855}
  {\path{doi:10.1002/pip.2855}}.

\bibitem{Beer66}
A.~C. Beer, R.~K. Willardson, Semiconductors and {Semimetals}: {Optical}
  properties of {III}-{V} compounds, Academic Press, 1966, google-Books-ID:
  s04AMQAACAAJ.

\bibitem{Kurtz90}
S.~R. Kurtz, P.~Faine, J.~M. Olson, Modeling of two‐junction,
  series‐connected tandem solar cells using top‐cell thickness as an
  adjustable parameter, Journal of Applied Physics 68~(4) (1990) 1890--1895.
\newblock \href {http://dx.doi.org/10.1063/1.347177}
  {\path{doi:10.1063/1.347177}}.

\bibitem{Kavanagh10}
K.~L. Kavanagh, Misfit dislocations in nanowire heterostructures, Semicond.
  Sci. Technol. 25~(2) (2010) 024006.
\newblock \href {http://dx.doi.org/10.1088/0268-1242/25/2/024006}
  {\path{doi:10.1088/0268-1242/25/2/024006}}.

\bibitem{Kayes05}
B.~M. Kayes, H.~A. Atwater, N.~S. Lewis, Comparison of the device physics
  principles of planar and radial p-n junction nanorod solar cells, Journal of
  Applied Physics 97~(11) (2005) 114302.
\newblock \href {http://dx.doi.org/10.1063/1.1901835}
  {\path{doi:10.1063/1.1901835}}.

\bibitem{LaPierre13}
R.~R. LaPierre, A.~C.~E. Chia, S.~J. Gibson, C.~M. Haapamaki, J.~Boulanger,
  R.~Yee, P.~Kuyanov, J.~Zhang, N.~Tajik, N.~Jewell, K.~M.~A. Rahman, {III–V}
  nanowire photovoltaics: {Review} of design for high efficiency, Phys. Status
  Solidi RRL 7~(10) (2013) 815--830.
\newblock \href {http://dx.doi.org/10.1002/pssr.201307109}
  {\path{doi:10.1002/pssr.201307109}}.

\bibitem{LaPierre11}
R.~R. LaPierre, Theoretical conversion efficiency of a two-junction {III}-{V}
  nanowire on {Si} solar cell, Journal of Applied Physics 110~(1) (2011)
  014310.
\newblock \href {http://dx.doi.org/10.1063/1.3603029}
  {\path{doi:10.1063/1.3603029}}.

\bibitem{LaPierre11a}
R.~R. LaPierre, Numerical model of current-voltage characteristics and
  efficiency of {GaAs} nanowire solar cells, Journal of Applied Physics 109~(3)
  (2011) 034311.
\newblock \href {http://dx.doi.org/10.1063/1.3544486}
  {\path{doi:10.1063/1.3544486}}.

\bibitem{Huang12}
N.~Huang, C.~Lin, M.~L. Povinelli, Limiting efficiencies of tandem solar cells
  consisting of {III}-{V} nanowire arrays on silicon, Journal of Applied
  Physics 112~(6) (2012) 064321.
\newblock \href {http://dx.doi.org/10.1063/1.4754317}
  {\path{doi:10.1063/1.4754317}}.

\bibitem{Bu13}
S.~Bu, X.~Li, L.~Wen, X.~Zeng, Y.~Zhao, W.~Wang, Y.~Wang, Optical and
  electrical simulations of two-junction {III}-{V} nanowires on {Si} solar
  cell, Applied Physics Letters 102~(3) (2013) 031106.
\newblock \href {http://dx.doi.org/10.1063/1.4788750}
  {\path{doi:10.1063/1.4788750}}.

\bibitem{Wang15}
Y.~Wang, Y.~Zhang, D.~Zhang, S.~He, X.~Li, Design {High}-{Efficiency}
  {III}–{V} {Nanowire}/{Si} {Two}-{Junction} {Solar} {Cell}, Nanoscale
  Research Letters 10~(1) (2015) 269.
\newblock \href {http://dx.doi.org/10.1186/s11671-015-0968-2}
  {\path{doi:10.1186/s11671-015-0968-2}}.

\bibitem{Chia13}
A.~C.~E. Chia, R.~R. LaPierre, Electrostatic model of radial pn junction
  nanowires, Journal of Applied Physics 114~(7) (2013) 074317.
\newblock \href {http://dx.doi.org/10.1063/1.4818958}
  {\path{doi:10.1063/1.4818958}}.

\bibitem{Li15b}
Z.~Li, Y.~C. Wenas, L.~Fu, S.~Mokkapati, H.~H. Tan, C.~Jagadish, Influence of
  {Electrical} {Design} on {Core}-{Shell} {GaAs} {Nanowire} {Array} {Solar}
  {Cells}, IEEE Journal of Photovoltaics 5~(3) (2015) 854--864.
\newblock \href {http://dx.doi.org/10.1109/JPHOTOV.2015.2405753}
  {\path{doi:10.1109/JPHOTOV.2015.2405753}}.

\bibitem{Moharam95}
M.~G. Moharam, T.~K. Gaylord, E.~B. Grann, D.~A. Pommet, Formulation for stable
  and efficient implementation of the rigorous coupled-wave analysis of binary
  gratings, J. Opt. Soc. Am. A, JOSAA 12~(5) (1995) 1068--1076.
\newblock \href {http://dx.doi.org/10.1364/JOSAA.12.001068}
  {\path{doi:10.1364/JOSAA.12.001068}}.

\bibitem{Bucci12}
D.~Bucci, B.~Martin, A.~Morand, Application of the three-dimensional aperiodic
  {Fourier} modal method using arc elements in curvilinear coordinates, J. Opt.
  Soc. Am. A, JOSAA 29~(3) (2012) 367--373.
\newblock \href {http://dx.doi.org/10.1364/JOSAA.29.000367}
  {\path{doi:10.1364/JOSAA.29.000367}}.

\bibitem{Michallon14}
J.~Michallon, D.~Bucci, A.~Morand, M.~Zanuccoli, V.~Consonni,
  A.~Kaminski-Cachopo, Light trapping in {ZnO} nanowire arrays covered with an
  absorbing shell for solar cells, Opt. Express, OE 22~(104) (2014)
  A1174--A1189.
\newblock \href {http://dx.doi.org/10.1364/OE.22.0A1174}
  {\path{doi:10.1364/OE.22.0A1174}}.

\bibitem{Songmuang16}
R.~Songmuang, L.~T.~T. Giang, J.~Bleuse, M.~Den~Hertog, Y.~M. Niquet, L.~S.
  Dang, H.~Mariette, Determination of the optimal shell thickness for
  self-catalyzed gaas/algaas core–shell nanowires on silicon, Nano Letters
  16~(6) (2016) 3426--3433.
\newblock \href {http://dx.doi.org/10.1021/acs.nanolett.5b03917}
  {\path{doi:10.1021/acs.nanolett.5b03917}}.

\bibitem{Utsumi98}
K.~Utsumi, O.~Matsunaga, T.~Takahata, Low resistivity {ITO} film prepared using
  the ultra high density {ITO} target, Thin Solid Films 334~(1–2) (1998)
  30--34.
\newblock \href {http://dx.doi.org/10.1016/S0040-6090(98)01111-0}
  {\path{doi:10.1016/S0040-6090(98)01111-0}}.

\bibitem{Konig14}
T.~A.~F. K\"{o}nig, P.~A. Ledin, J.~Kerszulis, M.~A. Mahmoud, M.~A. El-Sayed,
  J.~R. Reynolds, V.~V. Tsukruk, Electrically {Tunable} {Plasmonic} {Behavior}
  of {Nanocube}–{Polymer} {Nanomaterials} {Induced} by a {Redox}-{Active}
  {Electrochromic} {Polymer}, ACS Nano 8~(6) (2014) 6182--6192.
\newblock \href {http://dx.doi.org/10.1021/nn501601e}
  {\path{doi:10.1021/nn501601e}}.

\bibitem{Refractiveindex16}
Refractive index database, http://refractiveindex.info (accessed 05.05.2016.).

\bibitem{AM1.5D}
{Standard tables for reference solar spectral irradiances, ASTM G173-03 Tables,
  http://rredc.nrel.gov/solar/spectra/am1.5/ (accessed 20.01.2016.)}.

\bibitem{Sturmberg14}
B.~C.~P. Sturmberg, K.~B. Dossou, L.~C. Botten, A.~A. Asatryan, C.~G. Poulton,
  R.~C. McPhedran, C.~M. de~Sterke, Optimizing {Photovoltaic} {Charge}
  {Generation} of {Nanowire} {Arrays}: {A} {Simple} {Semi}-{Analytic}
  {Approach}, ACS Photonics 1~(8) (2014) 683--689.
\newblock \href {http://dx.doi.org/10.1021/ph500212y}
  {\path{doi:10.1021/ph500212y}}.

\bibitem{Huang12a}
N.~Huang, C.~Lin, M.~L. Povinelli, Broadband absorption of semiconductor
  nanowire arrays for photovoltaic applications, J. Opt. 14~(2) (2012) 024004.
\newblock \href {http://dx.doi.org/10.1088/2040-8978/14/2/024004}
  {\path{doi:10.1088/2040-8978/14/2/024004}}.

\bibitem{Anttu13}
N.~Anttu, H.~Q. Xu, Efficient light management in vertical nanowire arrays for
  photovoltaics, Opt. Express, OE 21~(103) (2013) A558--A575.
\newblock \href {http://dx.doi.org/10.1364/OE.21.00A558}
  {\path{doi:10.1364/OE.21.00A558}}.

\bibitem{Sturmberg12}
B.~C.~P. Sturmberg, K.~B. Dossou, L.~C. Botten, A.~A. Asatryan, C.~G. Poulton,
  R.~C. McPhedran, C.~M.~d. Sterke, Nanowire array photovoltaics: {Radial}
  disorder versus design for optimal efficiency, Applied Physics Letters
  101~(17) (2012) 173902.
\newblock \href {http://dx.doi.org/10.1063/1.4761957}
  {\path{doi:10.1063/1.4761957}}.

\bibitem{Foldyna13}
M.~Foldyna, L.~Yu, P.~Roca~i Cabarrocas, Theoretical short-circuit current
  density for different geometries and organizations of silicon nanowires in
  solar cells, Solar Energy Materials and Solar Cells 117 (2013) 645--651.
\newblock \href {http://dx.doi.org/10.1016/j.solmat.2012.10.014}
  {\path{doi:10.1016/j.solmat.2012.10.014}}.

\bibitem{Fountaine14a}
K.~T. Fountaine, C.~G. Kendall, H.~A. Atwater, Near-unity broadband absorption
  designs for semiconducting nanowire arrays via localized radial mode
  excitation, Opt. Express, OE 22~(103) (2014) A930--A940.
\newblock \href {http://dx.doi.org/10.1364/OE.22.00A930}
  {\path{doi:10.1364/OE.22.00A930}}.

\bibitem{Duan16}
Z.~Duan, M.~Li, T.~Mwenya, P.~Fu, Y.~Li, D.~Song, Effective light absorption
  and its enhancement factor for silicon nanowire-based solar cell, Appl. Opt.,
  AO 55~(1) (2016) 117--121.
\newblock \href {http://dx.doi.org/10.1364/AO.55.000117}
  {\path{doi:10.1364/AO.55.000117}}.

\bibitem{Li15}
Y.~Li, M.~Li, P.~Fu, R.~Li, D.~Song, C.~Shen, Y.~Zhao, A comparison of
  light-harvesting performance of silicon nanocones and nanowires for
  radial-junction solar cells, Scientific Reports 5 (2015) 11532.
\newblock \href {http://dx.doi.org/10.1038/srep11532}
  {\path{doi:10.1038/srep11532}}.

\bibitem{Hu13}
Y.~Hu, M.~Li, J.-J. He, R.~R. LaPierre, Current matching and efficiency
  optimization in a two-junction nanowire-on-silicon solar cell, Nanotechnology
  24~(6) (2013) 065402.
\newblock \href {http://dx.doi.org/10.1088/0957-4484/24/6/065402}
  {\path{doi:10.1088/0957-4484/24/6/065402}}.

\bibitem{Adachi05}
S.~Adachi, Properties of {Group}-{IV}, {III}-{V} and {II}-{VI}
  {Semiconductors}: {Adachi}/{Properties} of {Group}-{IV}, {III}-{V} and
  {II}-{VI} {Semiconductors}, John Wiley \& Sons, Ltd, Chichester, UK, 2005.

\bibitem{Sentaurus}
{Synopsys Inc. Sentaurus Device User Guide, Sentaurus TCAD, Version
  J-2014.09-SP1}.

\bibitem{Garnett10}
E.~Garnett, P.~Yang, Light {Trapping} in {Silicon} {Nanowire} {Solar} {Cells},
  Nano Lett. 10~(3) (2010) 1082--1087.
\newblock \href {http://dx.doi.org/10.1021/nl100161z}
  {\path{doi:10.1021/nl100161z}}.

\bibitem{Zhu10}
J.~Zhu, C.-M. Hsu, Z.~Yu, S.~Fan, Y.~Cui, Nanodome {Solar} {Cells} with
  {Efficient} {Light} {Management} and {Self}-{Cleaning}, Nano Lett. 10~(6)
  (2010) 1979--1984.
\newblock \href {http://dx.doi.org/10.1021/nl9034237}
  {\path{doi:10.1021/nl9034237}}.

\bibitem{Yao14}
M.~Yao, N.~Huang, S.~Cong, C.-Y. Chi, M.~A. Seyedi, Y.-T. Lin, Y.~Cao, M.~L.
  Povinelli, P.~D. Dapkus, C.~Zhou, {GaAs} {Nanowire} {Array} {Solar} {Cells}
  with {Axial} p–i–n {Junctions}, Nano Lett. 14~(6) (2014) 3293--3303.
\newblock \href {http://dx.doi.org/10.1021/nl500704r}
  {\path{doi:10.1021/nl500704r}}.

\bibitem{Chia12}
A.~C.~E. Chia, R.~R. LaPierre, Analytical model of surface depletion in {GaAs}
  nanowires, Journal of Applied Physics 112~(6) (2012) 063705.
\newblock \href {http://dx.doi.org/10.1063/1.4752873}
  {\path{doi:10.1063/1.4752873}}.

\bibitem{Neumann93}
H.~Neumann, S. {Adachi} (ed.). {Properties} of {Aluminium} {Gallium}
  {Arsenide}. ({EMIS} {Datareviews} {Series} {No}. 7). {INSPEC}; {The}
  {Institution} of {Electrical} {Engineers}, {London} 1993. 325 {S}., 58
  {Abb}., 93 {Tab}. {ISBN} 0 85296 558 3, {L} 95, Cryst. Res. Technol. 28~(6)
  (1993) 866--866.
\newblock \href {http://dx.doi.org/10.1002/crat.2170280617}
  {\path{doi:10.1002/crat.2170280617}}.

\bibitem{Pavesi94}
L.~Pavesi, M.~Guzzi, Photoluminescence of {AlxGa}1–{xAs} alloys, Journal of
  Applied Physics 75~(10) (1994) 4779--4842.
\newblock \href {http://dx.doi.org/10.1063/1.355769}
  {\path{doi:10.1063/1.355769}}.

\bibitem{Jiang12}
N.~Jiang, P.~Parkinson, Q.~Gao, S.~Breuer, H.~H. Tan, J.~Wong-Leung,
  C.~Jagadish, Long minority carrier lifetime in {Au}-catalyzed
  {GaAs}/{AlxGa}1-{xAs} core-shell nanowires, Applied Physics Letters 101~(2)
  (2012) 023111.
\newblock \href {http://dx.doi.org/10.1063/1.4735002}
  {\path{doi:10.1063/1.4735002}}.

\bibitem{Kuhn00}
R.~Kuhn, P.~Fath, E.~Bucher, Effects of pn-junctions bordering on surfaces
  investigated by means of 2d-modeling, in: Conference Record of the
  Twenty-Eighth IEEE Photovoltaic Specialists Conference - 2000 (Cat.
  No.00CH37036), 2000, pp. 116--119.

\bibitem{Kessler12}
M.~Kessler, T.~Ohrdes, P.~P. Altermatt, R.~Brendel, The effect of sample edge
  recombination on the averaged injection-dependent carrier lifetime in
  silicon, Journal of Applied Physics 111~(5) (2012) 054508.
\newblock \href {http://dx.doi.org/10.1063/1.3691230}
  {\path{doi:10.1063/1.3691230}}.

\bibitem{Bertrand17}
D.~Bertrand, S.~Manuel, M.~Pirot, A.~Kaminski-Cachopo, Y.~Veschetti, Modeling
  of edge losses in al-bsf silicon solar cells, IEEE Journal of Photovoltaics
  7~(1) (2017) 78--84.
\newblock \href {http://dx.doi.org/10.1109/JPHOTOV.2016.2618603}
  {\path{doi:10.1109/JPHOTOV.2016.2618603}}.

\end{thebibliography}
\bibliographystyle{elsarticle-num}

\end{document}